\documentclass{style}
\usepackage{graphicx}
\usepackage{color}
\usepackage{bm}
\usepackage{citesort}

\usepackage{amsmath}

\begin{document}

\title{Multidimensionally-constrained covariant density functional theories ---
nuclear shapes and potential energy surfaces}
\author{
  Shan-Gui Zhou$^{1,2,3,4}$\email{sgzhou@itp.ac.cn} \\
  \it $^{1}$Key Laboratory of Theoretical Physics, Institute of Theoretical Physics, \\
  \it Chinese Academy of Sciences, Beijing 100190, China \\
  \it $^{2}$School of Physics, University of Chinese Academy of Sciences, Beijing 100049, China \\
  \it $^{3}$Center of Theoretical Nuclear Physics, National Laboratory of Heavy Ion Accelerator, \\
  \it Lanzhou 730000, China \\ 
  \it $^{4}$Synergetic Innovation Center for Quantum Effects and Application, Hunan Normal University, \\
  \it Changsha, 410081, China
}
\pacs{21.60.Jz, 25.85.-w, 24.75.+i, 27.90.+b}
\maketitle

\begin{abstract}
The intrinsic nuclear shapes deviating from a sphere not only manifest 
themselves in nuclear collective states but also play important roles
in determining nuclear potential energy surfaces (PES's) and fission barriers.
In order to describe microscopically and self-consistently nuclear shapes
and PES's with as many shape degrees of freedom as possible included, 
we developed 
multidimensionally-constrained covariant density functional theories
(MDC-CDFTs).
In MDC-CDFTs, the axial symmetry and the reflection symmetry are both broken and
all deformations characterized by $\beta_{\lambda\mu}$ with even $\mu$ are considered.  
We have used the MDC-CDFTs to study PES's and fission barriers of actinides,
the non-axial octupole $Y_{32}$ correlations in $N = 150$ isotones
and shapes of hypernuclei.
In this Review we will give briefly the formalism of MDC-CDFTs
and present the applications to normal nuclei.
\end{abstract}


\section{Introduction}
\label{sec:intro}

The ``shape'' provides an intuitive understanding of the spatial distribution
of matter density in a quantum many-body system, such as
molecules \cite{Jahn1937_PRSA161-220,Simons2008_JChemPhysA112-6401},
atoms \cite{Ceraulo1991_PRA44-4145}, 
atomic nuclei~\cite{Bohr1998_Nucl_Structure_1,Bohr1998_Nucl_Structure_2,Ring1980} and
mesons \cite{Alexandrou2008_PRD78-094506}. 
In atomic nuclei, the occurrence of spontaneous symmetry breaking results in
various shapes which are associated with different spatial symmetries.
If the shape of an atomic nucleus deviates from a sphere, we call
it a deformed nucleus.
The nuclear deformation manifests itself in many observable phenomena,
including the small and large amplitude collective motions with 
the rotation and the fission as typical examples 
\cite{Frauendorf2001_RMP73-463,Nazarewicz2001_LNP581-102}.

\begin{figure}
\begin{center}
\resizebox{1.0\columnwidth}{!}{%
 \includegraphics{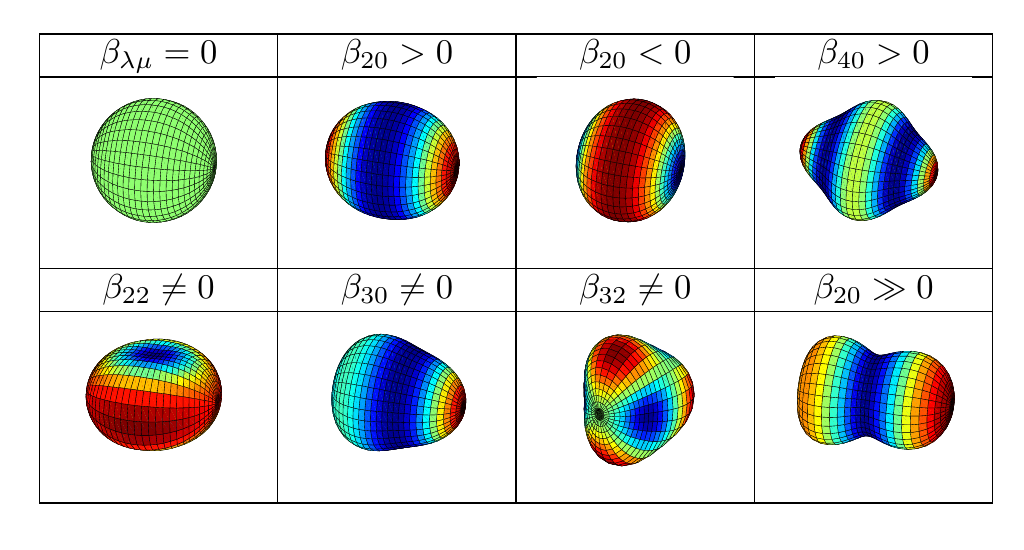} }
\end{center}
\caption{\label{Pic:shapes}(Color online)
Some typical nuclear shapes. 
In the first row (from left to right), 
a sphere ($\beta_{\lambda\mu} > 0$), 
a prolate spheroid ($\beta_{20} > 0$), 
an oblate spheroid ($\beta_{20} < 0$) and 
a hexadecapole shape ($\beta_{20} > 0$ and $\beta_{40} > 0$) are shown.
In the second row (from left to right), 
a triaxial ellipsoid ($\beta_{20} > 0$ and $\beta_{22} \ne 0$),
a reflection asymmetric octupole shape ($\beta_{30} \ne 0$), 
a tetrahedron [$\beta_{32} \ne 0$ and $\beta_{\lambda\mu}=0$ for all other $(\lambda\mu)$'s]
and a largely elongated and reflection asymmetric shape 
($\beta_{20} \gg 0$ and $\beta_{30} \ne 0$)
are shown. 
Taken from Ref.~\cite{Lu2012_PhD}.
}
\end{figure}

There are mainly two kinds of ways to parametrize 
the nuclear shape \cite{Brack1972_RMP44-320,Hasse1988}.
One of them is the two-center or two-center-like parametrization \cite{Mustafa1972_PRL28-1536,%
Swiatecki1981_PS24-113,Moeller2001_Nature409-785,Diaz-Torres2008_PRL101-122501,%
Sun2013_ChinPhysC37-014102}.
The other way is of one-center and to make a multipole expansion,
\begin{equation}
 R ( \theta, \varphi ) = 
 R_0 \left[  1 + 
           \sum_{\lambda=1}^{\infty} \sum_{\mu=-\lambda}^\lambda 
            \beta_{\lambda \mu}^* Y_{\lambda \mu} ( \theta, \varphi )    
     \right] ,
 \label{Eq:SurfaceDeformation}
\end{equation}
where $\beta_{\lambda \mu}$ is the deformation parameter. 
The way of multipole expansion is usually used in mean field calculations.
Several typical nuclear shapes are shown schematically in Fig.~\ref{Pic:shapes}.

Most of the nuclei have shapes similar as a spheroid and such shapes are
often described by $\beta_{20}$ --- an axial quadrupole deformation parameter.
The Nilsson perturbed-spheroid parameter $\epsilon_2$  
\cite{Nilsson1955_DMFM29-16,Nilsson1969_NPA131-1} was also
often adopted for numerical convenience.
It has been predicted for quite a long time that in some atomic nuclei, 
the non-axial quadrupole (triaxial) deformation $\beta_{22}$ (or $\gamma$ 
in the Hill-Wheeler coordinates) plays an important role.
A static triaxial shape manifests itself by the wobbling motion or
chiral doublet bands; both have been extensively studied from experimental and
theoretical sides \cite{Frauendorf1997_NPA617-131,Starosta2001_PRL86-971,%
Odegard2001_PRL86-5866,Chen2011_NSC2010,%
Meng2010_JPG37-064025,Meng2013_FPC8-55,Meng2014_IJMPE23-1430016}.
The existence of multiple chiral doublet (M$\chi$D) bands
in one nucleus was predicted by Meng et al. \cite{Meng2006_PRC73-037303} and
later multiple chiral doublet bands were 
observed \cite{Ayangeakaa2013_PRL110-172504,Lieder2014_PRL112-202502,%
Kuti2014_PRL113-032501,Tonev2014_PRL112-052501,Liu2016_PRL116-112501}.
Recently, collective Hamiltonians have been proposed to study chiral 
and wobbling modes \cite{Chen2013_PRC87-024314,Chen2014_PRC90-044306}.
It is desirable to construct similar Hamiltonians (the collective potential
and mass parameters) based on covariant density functionals.
In Ref.~\cite{Bengtsson1984_NPA415-189}, Bengtsson et al. proposed that 
one of the fingerprints of the triaxiality could be the low-spin signature
inversion which has been also a hot topic in low energy nuclear structure
studies \cite{Liu1995_PRC52-2514,Liu1996_PRC54-719,Zhou1996_JPG22-415,%
Riedinger1997_PPNP38-251,Liu1998_PRC58-1849}.
Moreover, the termination of rotational bands is connected with
the development of triaxial shapes in atomic nuclei \cite{Afanasjev1999_PR322-1}
and the triaxiality may also play a crucial role in superheavy nuclei 
\cite{Cwiok2005_Nature433-705}. 



The octupole shapes characterized by $\beta_{30}$ were predicted 
to be very pronounced in many nuclei, 
see Ref.~\cite{Butler1996_RMP68-349} for a review.
Low-lying negative-parity levels, connected with the ground state
bands via strong $E1$ transitions, in actinides and some
rare-earth nuclei are related to reflection asymmetric shapes 
with non-zero $\beta_{30}$ 
\cite{Shneidman2003_PRC67-014313,Shneidman2006_PRC74-034316,%
Wang2005_PRC72-024317,%
Yang2009_CPL26-082101,Robledo2011_PRC84-054302,Zhu2012_PRC85-014330,Zhu2012_NSC2012-348,%
Nomura2013_PRC88-021303R,Nomura2014_PRC89-024312,%
Nomura2015_PRC92-014312}.
The evidence for a strong octupole deformation has been shown
in an experimental study of electric octupole transition strengths  
in $^{220}$Rn and $^{224}$Ra \cite{Gaffney2013_Nature497-199}
and in $^{144}\mathrm{Ba}$ \cite{Bucher2016_PRL116-112503}.
It was also revealed that the octupole deformation may explain 
the Ra puzzle \cite{Yu2012_arXiv1211.0601}:   
The residual proton-neutron interactions for Ra isotopes show 
an anomalous enhancement around $N=135$ 
\cite{Brenner2006_PRC73-034315,Neidherr2009_PRL102-112501}.
Recently, Liu et al. identified multiple chiral doublet bands in $^{78}$Br 
and observed octupole correlations between them \cite{Liu2016_PRL116-112501}.
This raises an interesting topic on possible existence of 
nuclear chirality-parity quartet bands.
The occurrence of the non-axial octupole $\beta_{32}$ deformation 
has also been investigated
\cite{Hamamoto1991_ZPD21-163,Skalski1991_PRC43-140,Li1994_PRC49-R1250,%
Takami1998_PLB431-242,%
Yamagami2001_NPA693-579,%
Dudek2002_PRL88-252502,Dudek2006_PRL97-072501,%
Olbratowski2006_IJMPE15-333,%
Zberecki2006_PRC74-051302R,%
Dudek2010_JPG37-064032}.
More discussions about the $\beta_{32}$ deformation will be given
in Section~\ref{sec:Y32}.

Nuclear deformations corresponding to higher-order terms with $\lambda>3$ in 
Eq.~(\ref{Eq:SurfaceDeformation}) are important to different extents.
The hexadecapole deformation, $\beta_{40}$ or $\epsilon_{4}$,
has been included in deformed mean field potentials since 1960s, 
see, e.g., Ref.~\cite{Nilsson1969_NPA131-1}.
The effects of the $\beta_{60}$ or $\epsilon_6$ deformation on the
alignment of angular momentum and dynamic moment of inertia ${\cal J}^{(2)}$ 
in some superheavy nuclei 
were recently revealed \cite{Liu2012_PRC86-011301R,Zhang2013_PRC87-054308}.

These various deformations are also very relevant to large amplitude nuclear collective 
motions, e.g., the fission.
Since the nuclear fission is discovered \cite{Hahn1939_Naturwissenschaften27-11}, 
the description of the fission process has been a difficult and challenging task.
The fission dynamics are mostly determined 
by the barriers which prohibit the dissolving of the nucleus \cite{Bohr1939_PR056-426}.
Since the barrier penetration is involved in the fission process, 
both the height and the width of the fission barrier are very crucial 
in calculating fission probability 
\cite{Zubov2009_PPN40-847,Xia2011_SciChinaPMA54S1-109}.
In particular, 
one of the forefronts of modern nuclear science is to explore
the existence of charge and mass limits of atomic nuclei, i.e., 
to study superheavy nuclei 
\cite{Hofmann2000_RMP72-733,%
Oganessian2007_JPG34-R165,Oganessian2010_PRL104-142502,%
Morita2004_JPSJ73-2593,Morita2012_JPSJ81-103201,%
Zhang2012_CPL29-012502}.
In order to describe the synthesis
mechanism of superheavy nuclei via heavy ion fusion reactions,
it becomes more desirable to have reliable descriptions for
fission barriers for superheavy nuclei \cite{Moeller2009_PRC79-064304,%
Peter2004_EPJA22-271,Peter2004_NPA734-192,%
Pei2009_PRL102-192501,%
Antonenko1995_PRC51-2635,Adamian1997_NPA618-176,%
Liu2007_PRC76-034604,Liu2013_PRC87-034616,%
Zubov2009_PPN40-847,Xia2011_SciChinaPMA54S1-109,Liu2011_PRC84-031602R,%
Gan2011_SciChinaPMA54S1-61,%
Wang2011_PRC84-061601R,Zanganeh2012_PRC85-034601,%
Walker2012_JPG39-105106,Wang2012_PRC85-041601R}.
Until now many popular nuclear structure models have been employed to calculate 
nuclear fission barriers, including the macroscopic-microscopic 
model \cite{Moeller2001_Nature409-785,%
Moeller2004_PRL92-072501,Moeller2009_PRC79-064304,%
Pomorski2003_PRC67-044316,Ivanyuk2009_PRC79-054327,%
Kowal2010_PRC82-014303,%
Royer2012_PRC86-044326,Xie2014_SciChinaPMA57-189},
the extended Thomas-Fermi plus Strutinsky integral 
method \cite{Mamdouh1998_NPA644-389,Mamdouh2001_NPA679-337}, 
the non-relativistic mean field models 
(Hartree-Fock or Hartree-Fock-Bogoliubov method with the Skyrme 
force~\cite{%
Burvenich2004_PRC69-014307,%
Bonneau2004_EPJA21-391,%
Samyn2005_PRC72-044316,Goriely2007_PRC75-064312,%
Skalski2007_PRC76-044603,%
Minato2009_NPA831-150,%
Jachimowicz2011_PRC83-054302,%
Kortelainen2012_PRC85-024304,%
McDonnell2012_arXiv1301.7587,%
Staszczak2013_PRC87-024320,%
Schunck2014_PRC90-054305,Schunck2015_PRC91-034327}
and the Gogny force \cite{Egido2000_PRL85-1198,Warda2012_PRC86-014322}),
and the covariant density functional theory 
\cite{Blum1994_PLB323-262,%
Zhang2003_CPL20-1694,%
Burvenich2004_PRC69-014307,%
Lu2006_CPL23-2940,%
Li2010_PRC81-064321,%
Abusara2010_PRC82-044303,Abusara2012_PRC85-024314,%
Lu2012_PRC85-011301R,Lu2012_PhD,Lu2012_EPJWoC38-05003,Zhao2012_PRC86-057304,%
Prassa2012_PRC86-024317,Afanasjev2012_IJMPE21-1250025,%
Afanasjev2013_ICFN5-303,Afanasjev2013_EPJWoC62-03003,%
Lu2014_PRC89-014323,Lu2014_PS89-054028,Lu2014_JPCS492-012014,%
Zhao2015_PRC91-014321}.
These models show similar trend of the barrier heights but 
the quantitative disagreement is rather large for some nuclei.

The axial quadrupole deformation $\beta_{20}$ 
characterizes the elongation of a nucleus and
is the most relevant shape for the nuclear fission.
Besides $\beta_{20}$ 
and $\beta_{40}$ which is connected with the neck formed during fission, 
many other shapes are indispensable  
for studying properties of the fission barrier and for locating 
the static fission path. 
Let us take actinides as examples.
Due to shell effects,
a double-humped fission barrier appears in actinide nuclei 
\cite{Brack1972_RMP44-320}.
There may also appear a third minimum in the PES's of light actinide nuclei
\cite{Bengtsson1987_NPA473-77,%
Thirolf2002_PPNP49-325,Royer2012_PRC86-044326,%
Csige2012_PRC85-054306,Kowal2012_PRC85-061302R,Jachimowicz2013_PRC87-044308}.
In 1970s it was found in calculations with macroscopic-microscopic models
that the height of inner fission barrier can be reduced by the triaxial distortion 
\cite{Pashkevich1969_NPA133-400,%
Moeller1970_PLB31-283,Randrup1976_PRC13-229}
and that of the outer one can be lowered by the reflection asymmetric (RA) 
distortion \cite{Ledergerber1973_NPA207-1}.
The importance of the triaxial and octupole distortions were later also confirmed 
both in non-relativistic \cite{Girod1983_PRC27-2317} 
and relativistic \cite{Rutz1995_NPA590-680,Abusara2010_PRC82-044303}
mean field calculations. 




From the above discussions, it is clear that many deformations 
are crucial in the ground state, low-lying states, potential energy
surfaces and fission properties of atomic nuclei.
It is thus very desirable to develop self-consistent theoretical models 
in which as many deformations as possible are considered.
This is especially true for studying atomic nuclei in unknown regions,
including nuclei close to proton and neutron drip lines, 
superheavy nuclei, et al.

Nowadays, one of the most successful self-consistent approaches is
the covariant density functional theory (CDFT). 
The CDFT has been used to describe 
both properties of ground states and excited states of nuclei in the whole
nuclear chart ranging from light to superheavy regions 
\cite{Serot1986_ANP16-1,Reinhard1989_RPP52-439,%
Ring1996_PPNP37-193,Bender2003_RMP75-121,Vretenar2005_PR409-101,%
Meng2006_PPNP57-470,Paar2007_RPP70-691,Niksic2011_PPNP66-519,%
Liang2015_PR570-1,Meng2015_JPG42-093101,Meng2016_WorldSci},
including PES's and fission barriers of heavy and superheavy nuclei
\cite{Bender2003_RMP75-121,Abusara2010_PRC82-044303,%
Lu2012_PRC85-011301R,Lu2012_PhD,Lu2012_EPJWoC38-05003,Zhao2012_PRC86-057304,%
Abusara2012_PRC85-024314,Prassa2012_PRC86-024317,Afanasjev2012_IJMPE21-1250025,%
Afanasjev2013_ICFN5-303,Afanasjev2013_EPJWoC62-03003,%
Lu2014_PRC89-014323,Lu2014_PS89-054028,Lu2014_JPCS492-012014,%
Zhao2015_PRC91-014321,Lu2016_RDFNS-171}.
In the last years, 
we have developed multidimensionally-constrained covariant density functional theories (MDC-CDFTs) 
in which the reflection symmetry and the axial symmetry are both broken
\cite{Lu2012_PhD,Lu2012_PRC85-011301R,Lu2014_PRC89-014323,Zhao2016_in-prep}.
In MDC-CDFTs, 
all $\beta_{\lambda\mu}$'s with even $\mu$, 
i.e., $\beta_{20}$, $\beta_{22}$, $\beta_{30}$, $\beta_{32}$, $\beta_{40}$, $\cdots$, 
are included self-consistently.
In CDFTs, either the meson exchange (ME) or 
point-coupling (PC) interactions can be used. 
Furthermore, the couplings can be non-linear (NL) or density-dependent (DD). 
All these four types of relativistic density functionals, i.e., 
NL-ME, DD-ME, NL-PC and DD-PC, were implemented in MDC-CDFTs.
The pairing correlations are treated with either the Bogoliubov approach 
or the BCS method.
The MDC-CDFTs have been applied to the study of fission barriers and PES's 
of actinide and trans-actinide nuclei 
\cite{Lu2012_PRC85-011301R,Lu2012_EPJWoC38-05003,Lu2014_PRC89-014323,%
Lu2014_JPCS492-012014,Lu2014_PS89-054028,Zhao2015_PRC92-064315,Zhao2016_in-prep2}, 
the non-axial octupole $Y_{32}$ effects in four $N = 150$ isotones 
\cite{Zhao2012_PRC86-057304},
the third minima and triple-humped barriers in light actinides \cite{Zhao2015_PRC91-014321}
and shapes of hypernuclei \cite{Lu2011_PRC84-014328,Lu2014_PRC89-044307}.

We present a Review of the MDC-CDFTs and the applications of MDC-CDFTs on
actinide and trans-actinide nuclei.
In Section~\ref{sec:formalism}, we briefly give the formalism of the MDC-CDFTs.
The study of PES's and fission barriers of actinide nuclei and 
the third barriers in light actinides will be discussed in Section~\ref{sec:PES}.
In Section~\ref{sec:Y32}, we present results of non-axial octupole $Y_{32}$ effects 
in $N = 150$ isotones.
A summary and some perspectives are given in Section~\ref{sec:summary}.


\section{Formalism}
\label{sec:formalism}

In the covariant density functional theory, 
a nucleus is described as a composite of nucleons which interact
either through exchanges of mesons and photons
or by point-couplings 
\cite{Serot1986_ANP16-1,Reinhard1989_RPP52-439,%
Ring1996_PPNP37-193,Bender2003_RMP75-121,Vretenar2005_PR409-101,%
Meng2006_PPNP57-470,Paar2007_RPP70-691,Niksic2011_PPNP66-519,%
Liang2015_PR570-1,Meng2015_JPG42-093101,Meng2016_WorldSci}.
The nonlinear coupling terms 
\cite{Boguta1977_NPA292-413,Brockmann1992_PRL68-3408,Sugahara1994_NPA579-557}
or
the density-dependent coupling constants \cite{Fuchs1995_PRC52-3043,%
Niksic2002_PRC66-024306}
were introduced in order to give a proper description for nuclear matter
(e.g., the saturation properties).
Accordingly, there are four kinds of relativistic density functionals.
In this Section, we take the nonlinear point-couplings (NL-PC) covariant 
density functional theory as an example and present briefly the formalism
of MDC-CDFT.
The time-reversal symmetry is assumed for the nuclei in question. 

The NL-PC Lagrangian reads,
\begin{equation}
 \mathcal{L} = \bar{\psi}(i\gamma_{\mu}\partial^{\mu}-M)\psi
              -\mathcal{L}_{{\rm lin}}
              -\mathcal{L}_{{\rm nl}}
              -\mathcal{L}_{{\rm der}}
              -\mathcal{L}_{{\rm Cou}},
\end{equation}
where the linear coupling term, the nonlinear coupling term, the derivative coupling term,
and the Coulomb term are given as follows,
\begin{eqnarray}
 \mathcal{L}_{{\rm lin}} & = & \frac{1}{2} \alpha_{S} \rho_{S}^{2}
                              +\frac{1}{2} \alpha_{V} \rho_{V}^{2}
                              +\frac{1}{2} \alpha_{TS} \vec{\rho}_{TS}^{2}
                              +\frac{1}{2} \alpha_{TV} \vec{\rho}_{TV}^{2} ,
 \nonumber \\
 \mathcal{L}_{{\rm nl}}  & = & \frac{1}{3} \beta_{S} \rho_{S}^{3}
                              +\frac{1}{4} \gamma_{S}\rho_{S}^{4}
                              +\frac{1}{4} \gamma_{V}[\rho_{V}^{2}]^{2} ,
 \nonumber \\
 \mathcal{L}_{{\rm der}} & = & \frac{1}{2} \delta_{S}[\partial_{\nu}\rho_{S}]^{2}
                              +\frac{1}{2} \delta_{V}[\partial_{\nu}\rho_{V}]^{2}
                              +\frac{1}{2} \delta_{TS}[\partial_{\nu}\vec{\rho}_{TS}]^{2}
 \nonumber \\
 &  & \mbox{}                 +\frac{1}{2} \delta_{TV}[\partial_{\nu}\vec{\rho}_{TV}]^{2} ,
 \nonumber \\
 \mathcal{L}_{{\rm Cou}} & = & \frac{1}{4} F^{\mu\nu} F_{\mu\nu}
                             +e\frac{1-\tau_{3}}{2} A_{0} \rho_{V} .
\label{eq:lagrangian}
\end{eqnarray}
$M$ is the nucleon mass and $\alpha_{S}$, $\alpha_{V}$, $\alpha_{TS}$,
$\alpha_{TV}$, $\beta_{S}$, $\gamma_{S}$, $\gamma_{V}$, $\delta_{S}$,
$\delta_{V}$, $\delta_{TS}$ and $\delta_{TV}$ are coupling constants
corresponding to different channels. 
$\rho_{S}$ and $\vec{\rho}_{TS}$ are the isoscalar and isovector densities;
$\rho_{V}$ and $\vec{\rho}_{TV}$ are the time-like components of isoscalar 
and isovector currents.
These densities and currents read,
\begin{equation}
 \rho  _{S} = \bar{\psi}\psi , \ 
 \vec{\rho}_{TS} = \bar{\psi}\vec{\tau}\psi , \ 
      \rho_{V} = \bar{\psi} \gamma^{0} \psi , \ 
 \vec{\rho}_{TV} = \bar{\psi} \vec{\tau} \gamma^{0} \psi.
  \label{eq:densities}
\end{equation}

With the Slater determinants used as trial wave functions,
applying the mean field approximation and the no-sea approximation, 
one can derive the Dirac equation,
\begin{equation}
 \hat{h}\psi_{k}(\bm{r}) = \epsilon_{k} \psi_{k}(\bm{r}) ,
 \label{eq:Diracequation}
\end{equation}
where
\begin{equation}
 \hat{h} = \bm{\alpha} \cdot \bm{p}
         + \beta \left[ M+S(\bm{r}) \right]
         + V(\bm{r}) ,
 \label{eq:dirac}
\end{equation}
is the single-particle Dirac Hamiltonian which consists of two
potentials, the scalar potential and the vector potential,
\begin{eqnarray}
 S & = &
  \alpha_{S} \rho_{S}     + \alpha_{TS} \vec{\rho}_{TS} \cdot \vec{\tau}
 + \beta_{S} \rho_{S}^{2} +  \gamma_{S} \rho_{S}^{3}
 \nonumber \\
 &  & \mbox{}
 +\delta_{S} \triangle \rho_{S} + \delta_{TS} \triangle \vec{\rho}_{TS} \cdot \vec{\tau}
 ,
 \\
%
V & = &
   \alpha_{V} \rho_{V} + \alpha_{TV} \vec{\rho}_{TV} \cdot \vec{\tau}
 + \gamma_{V} \rho_{V}^{2} \rho_{V}
 \nonumber \\
 &  & \mbox{}
 + \delta_{V} \triangle \rho_{V} + \delta_{TV} \triangle \vec{\rho}_{TV} \cdot \vec{\tau}
 .
\label{eq:potential}
\end{eqnarray}

We can solve the RMF equations either in the coordinate $r$ space or
in a complete basis. 
These equations and various extensions were already solved in $r$ space
for spherical nuclei, including the RMF equations 
\cite{Horowitz1981_NPA368-503,Horowitz1984_PLB140-86},
the relativistic Hartree-Bogoliubov (RHB) equations 
\cite{Meng1996_PRL77-3963,Meng1998_PRL80-460,Meng1998_NPA635-3,Poschl1997_PRL79-3841,%
Poschl1997_CPC101-75,Poschl1997_CPC103-217}, 
the relativistic Hartree-Fock (RHF) equations 
\cite{Long2006_PLB640-150,Long2006_PhD,%
Long2006_PLB639-242,Long2007_PRC76-034314,Long2008_EPL82-12001,Long2009_PLB680-428}, 
and the random phase approximation (RPA) based on the RMF and RHF
models \cite{Liang2008_PRL101-122502,Liang2009_PRC79-064316,Liang2010_PhD}.
For deformed nuclei, however, it is very difficult to do so because, in addition
to conventional complications related to two-dimensional or three-dimensional spatial
lattice techniques 
\cite{Pei2008_PRC78-064306,Pei2011_PRC84-024311,Pei2013_PRC87-051302R,Pei2014_PRC90-024317}, 
one also has to deal with problems of the variational collapse 
\footnote{The variational collapse is also a problem one has to solve in
the basis expansion method, see, e.g., Section IIIA and 
Fig.~3 of Ref.~\cite{Lu2014_PRC89-014323} and references therein.} 
and fermion doubling in the relativistic framework.
Nevertheless, the RMF equations have been solved 
on a grid in cylindrical coordinates 
for axially deformed nuclei 
\cite{Lee1986_PRL57-2916,Lee1987_PRL59-1171,Rutz1995_NPA590-680}.
Recently, Tanimura et al. developed a novel method to solve the RMF equations 
in three-dimensional lattice \cite{Tanimura2015_PTEP2015-073D01}. 

It is more convenient to solve the RMF equations for deformed nuclei
by expanding the single-particle Dirac wave functions in terms of a complete basis.
Up to now, several bases have been proposed and used widely when
solving the RMF equations, including
the axially deformed harmonic oscillator basis 
\cite{Gambhir1990_APNY198-132,Ring1997_CPC105-77}, 
the three-dimensional Cartesian harmonic oscillator basis 
\cite{Koepf1988_PLB212-397,Koepf1989_NPA493-61,Afanasjev2000_NPA676-196},
the transformed harmonic oscillator basis \cite{Stoitsov1998_PRC58-2092,Stoitsov1998_PRC58-2086,%
Stoitsov2000_PRC61-034311,Stoitsov2003_PRC68-054312,%
Stoitsov2013_CPC184-1592},
the P\"oschl-Teller-Ginocchio basis \cite{Stoitsov2008_PRC77-054301},
and the Woods-Saxon basis \cite{Zhou2003_PRC68-034323,Zhou2006_AIPCP865-90,%
Zhou2010_PRC82-011301R,Long2010_PRC81-024308,Li2012_PRC85-024312,Li2012_CPL29-042101,%
Chen2012_PRC85-067301,Niu2013_PLB723-172}.
Furthermore, Geng et al. have developed 
a reflection asymmetric relativistic mean field (RAS-RMF) model
in a basis generated by a two-center harmonic oscillator potential \cite{Geng2007_CPL24-1865};
this model has been 
used to study even-even $^{146-156}$Sm
in which the important role of the octupole deformation on shape phase transitions
was found \cite{Zhang2010_PRC81-034302}.
In the MDC-CDFTs, we adopted
an axially deformed harmonic oscillator (ADHO) basis 
\cite{Gambhir1990_APNY198-132,Ring1997_CPC105-77}.
The axially deformed harmonic oscillator basis consists of the eigenstates 
of the Schr\"odinger equation,
\begin{eqnarray}
 \left[-\frac{\hbar^{2}}{2M}\nabla^{2}+V_{B}(z,\rho)\right]\Phi_{\alpha}(\bm{r}\sigma)
 & = & E_{\alpha}\Phi_{\alpha}(\bm{r}\sigma) ,
 \label{eq:BasSchrodinger-1}
\end{eqnarray}
where $\bm{r} = (z,\rho)$ with $\rho=\sqrt{x^2+y^2}$ and
\begin{equation}
 V_{B}(z,\rho) = \frac{1}{2} M ( \omega_{\rho}^{2} \rho^{2} + \omega_{z}^{2} z^{2})
 ,
\end{equation}
is the axially deformed harmonic oscillator potential with 
$\omega_{\rho}$ ($\omega_{z}$) being 
the oscillator frequencies perpendicular to (along) the symmetry axis.
The solution of Eq.~(\ref{eq:BasSchrodinger-1}) is obtained as
\begin{equation}
 \Phi_{\alpha}(\bm{r}\sigma)
 =
 C_{\alpha} \phi_{n_{z}}(z) R_{n_{\rho}}^{m_{l}}(\rho)
 \frac{1}{\sqrt{2\pi}} e^{im_{l}\varphi}
 \chi_{m_s}(\sigma).
 \label{eq:HO}
\end{equation}
A complex number $C_{\alpha}$ is introduced for convenience. 
$\chi_{s_{z}}$ is a two-component spinor. 
$\phi_{n_{z}}(z)$ and $R_{n_{\rho}}^{m_{l}}(\rho)$ are the harmonic oscillator wave functions,
\begin{eqnarray}
 \phi_{n_{z}}(z) & = &
 \frac{1}{\sqrt{b_{z}}} \frac{1}{\pi^{{1}/{4}} \sqrt{2^{n_{z}}n_{z}!}}
  H_{n_{z}}\left(\frac{z}{b_{z}}\right)
  e^{-\frac{z^{2}}{2b_{z}}} ,
  \\
  R_{n_{\rho}}^{m_{l}}(\rho) & = &
  \frac{1}{b_{\rho}} \sqrt{\frac{2n_{\rho}!}{(n_{\rho}+|m_{l}|)!}}
  \left( \frac{\rho}{b_{\rho}} \right)^{|m_{l}|}
  \nonumber \\
  & & \mbox{\hspace{2cm}} \times
  L_{n_{\rho}}^{|m_{l}|}
  \left( \frac{\rho^{2}}{b_{\rho}^{2}} \right)
  e^{ -\frac{\rho^{2}}{2b_{\rho}^{2}} }
  .
\end{eqnarray}
Oscillator lengths $b_{z}$ and $b_{\rho}$ are related to the frequencies,
$b_{z}=1/\sqrt{M\omega_{z}}$ and $b_{\rho}=1/\sqrt{M\omega_{\rho}}$.
The corresponding eigenenergy
$E_{\alpha}=\omega_{\rho}(2n_{\rho}+|m_{l}|+1)+\omega_{z}(n_{z}+{1}/{2})$
and the major quantum number $N_{\alpha}=2n_{\rho}+|m_{l}|+n_{z}$.

These basis states
are eigenstates of 
$\hat{j}_{z}$ with eigenvalues $K_\alpha=m_{l}+m_{s}$.
We define the time-reversal state of $\Phi_{\alpha}(\bm{r}\sigma)$ as
$\Phi_{\bar{\alpha}}(\bm{r}\sigma)=\mathcal{T}\Phi_{\alpha}(\bm{r}\sigma)$
with the time-reversal operator $\mathcal{T}=i\sigma_{y}\hat{K}$ 
where 
$\hat{K}$ is the complex conjugation operator.
$K_{\bar{\alpha}}=-K_{\alpha}$ and $\pi_{\bar{\alpha}}=\pi_{\alpha}$
where $\pi_\alpha = \pm 1$ is the parity.
The deformation of the basis $\beta_{{\rm basis}}$ is defined through the relations
$\omega_{z}=\omega_{0}\exp\left(-\sqrt{{5}/{4\pi}}\beta_{{\rm basis}}\right)$
and $\omega_{\rho}=\omega_{0}\exp\left(\sqrt{{5}/{16\pi}}\beta_{{\rm basis}}\right)$,
where $\omega_{0}=(\omega_{z}\omega_{\rho}^{2})^{{1}/{3}}$ is the frequency of
the corresponding spherical oscillator.

When solving the RMF equations, the Dirac spinor with four components 
is expanded in terms of the complete basis $\{ \Phi_{\alpha}(\bm{r}\sigma) \}$ as,
\begin{equation}
 \psi_{i}(\bm{r}\sigma) =
 \left(
  \begin{array}{c}
   \sum_{\alpha}f_{i}^{\alpha} \Phi_{\alpha}(\bm{r}\sigma) \\
   \sum_{\alpha}g_{i}^{\alpha} \Phi_{\alpha}(\bm{r}\sigma)
  \end{array}
 \right),
\label{eq:spwaveexpansion}
\end{equation}
where 
$\alpha=\{n_{z},n_{\rho},m_{l},m_{s}\}$ and $f_{i}^{\alpha}$ and
$g_{i}^{\alpha}$ are the expansion coefficients.
A truncation has to be introduced in summations of Eq.~(\ref{eq:spwaveexpansion})
in practical calculations.
Following Ref.~\cite{Warda2002_PRC66-014310},
for the upper (large) component of the Dirac spinor, the basis states with 
$[ n_{z}/f_{z}+(2n_{\rho}+|m_l|)/f_{\rho} ] \le N_f$
are included in the summation where
$f_{z}=\max(b_{z}/b_{0},1)$ and $f_{\rho}=\max(b_{\rho}/b_{0},1)$ 
with $b_{0} = 1/\sqrt{M\omega_0}$.
For the expansion of the lower (small) component of the Dirac spinor, 
the truncation of $N_g = N_f+1$ is made.


When neither 
the spatial reflection symmetry nor the axial symmetry is kept, 
basis states with different \{$K,\pi$\} should be included
in the expansion in Eq.~(\ref{eq:spwaveexpansion}).
Thus the Hamiltonian matrix becomes much larger than 
axial or reflection symmetric cases.
Nevertheless, 
the simplex operator $\hat{S}=ie^{-i\pi \hat{j}_{z}}$
can still be used to
keep the Hamiltonian matrix to be block-diagonal.
$\hat{S}$ is Hermitian with $\hat{S}^{2}=1$ for a fermionic system. 
The basis states can be labelled with the good quantum number 
$S$ (the eigenvalue of $\hat{S}$),
$\hat{S}\Phi_{\alpha}=S\Phi_{\alpha}=(-1)^{K_{\alpha}-{1}/{2}}\Phi_{\alpha}$.
The whole Hilbert space can be divided into two subspaces:
One subspace with $S=+1$ consists of states $\Phi_{\alpha}$ with
$K_{\alpha}=+{1}/{2}$, $-{3}/{2}$, $+{5}/{2}$, $-{7}/{2},\ \dots$ and 
the other subspace with $S=-1$ 
consists of the corresponding time-reversal states 
with $K_{\alpha}=-{1}/{2}$, $+{3}/{2}$, $-{5}/{2}$, $+{7}/{2},\ \dots$.
Since the time-reversal symmetry is assumed, 
we can include in the expansion only the basis states with $S=+1$ for which 
we set $C_{\alpha}=1$ when expanding the large component and $C_{\alpha}=i$ 
for the small one [cf. Eq.~(\ref{eq:HO})].
We can then construct the Hamiltonian matrix with $S=+1$ and diagonalize it
to obtain the single-particle energies and wave functions. 
The Hamiltonian matrix with $S=-1$ can be obtained by applying
the time-reversal operation on that with $S=+1$.

In the MDC-CDFTs, we make Fourier series expansions for 
the densities in Eq.~(\ref{eq:densities}) and the two potentials $V(\bm{r})$ and $S(\bm{r})$,
\begin{equation}
 f(\rho,\varphi,z)
 = \sum_{\mu=-\infty}^{\infty} f_{\mu}(\rho,z)
   \frac{1}{\sqrt{2\pi}} \exp(i\mu\varphi)
 .
 \label{eq:potentialexpansion}
\end{equation}
As discussed in Refs.~\cite{Dobaczewski2000_PRC62-014310,Dobaczewski2000_PRC62-014311}, 
point-group symmetries have been widely used in mean field theories for atomic nuclei. 
In MDC-CDFTs, we assume that the nuclear potentials and densities are 
invariant under the following operations: 
the reflection with respect to the $y$-$z$ plane ($\hat{S}_x$), 
the reflection with respect to the $x$-$z$ plane ($\hat{S}_y$) and 
the rotation of 180$^\circ$ with respect to the $z$ axis ($\hat{S}$), i.e.,
\begin{eqnarray}
 \hat{S}_x \phi(x,y,z)  &=&  \phi(-x,y,z) , \\ 
 \hat{S}_y \phi(x,y,z)  &=&  \phi(x,-y,z) , \\ 
 \hat{S}   \phi(x,y,z)  &=&  \phi(-x,-y,z) .
\end{eqnarray}
These three operations and the identity $\hat{I}$ form the $V_4$ group 
\footnote{The $V_4$ (or $V$) group is the Klein's four-group 
(Vierergruppe in German) or Klein's group 
\cite{Armstrong1988_Groups-Symmetry,Ramond2010_GroupTheory}. 
It consists of four elements: $\{I, \sigma_1, \sigma_2, \sigma_3\}$ 
where $I$ is the identity, 
$\sigma^2_i = I$ and $\sigma_i \sigma_j = \sigma_k$ when $i \ne j \ne k$.}. 
Under the $V_4$ symmetry, 
\begin{equation}
 \hat{S}_x Y_{\lambda\mu} =        Y_{\lambda\bar\mu} ,\ 
 \hat{S}_y Y_{\lambda\mu} = (-1)^\mu Y_{\lambda\bar\mu} ,\ 
 \hat{S}  Y_{\lambda\mu} = (-1)^\mu Y_{\lambda\mu} .
\end{equation}
Applying these operations on Eq.~(\ref{Eq:SurfaceDeformation}), one gets,
\begin{equation}
 \beta_{\lambda\mu} =        \beta_{\lambda\bar\mu} 
 = (-1)^\mu \beta_{\lambda\bar\mu} .
\end{equation}
Therefore $\beta_{\lambda\mu} = 0$ when $\mu$ is an odd number.
Similarly, in Eq.~(\ref{eq:potentialexpansion}),
$f_{\mu}=f_{\mu}^{*}=f_{\bar{\mu}}$ and $f_{n}=0$ for odd $n$ 
and the expansion given in Eq.~(\ref{eq:potentialexpansion}) is then reduced as
\begin{equation}
 f(\rho,\varphi,z) =
  f_{0}(\rho,z) \frac{1}{\sqrt{2\pi}}
+ \sum_{n=1}^{\infty} f_{n}(\rho,z) \frac{1}{\sqrt{\pi}} \cos(2n\varphi),
\end{equation}
where
\begin{eqnarray}
 f_{0}(\rho_{,}z) & = & \frac{1}{\sqrt{2\pi}}\int_{0}^{2\pi}
                         d\varphi f(\rho,\varphi,z) ,
 \label{eq:potentialexpansion_0}
 \\
 f_{n}(\rho,z)    & = & \frac{1}{\sqrt{ \pi}}\int_{0}^{2\pi}
                         d\varphi f(\rho,\varphi,z)\cos(2n\varphi) ,
 \label{eq:potentialexpansion_n}
\end{eqnarray}
are all real functions of $\rho$ and $z$.
For the details about how to calculate the matrix elements and how to
deal with the densities and their derivatives, please refer to 
Refs.~\cite{Lu2014_PRC89-014323,Zhao2016_in-prep}.

The pairing correlations are crucial in open shell nuclei.
Karatzikos et al. have shown that fission barriers are influenced 
by the pairing force very much \cite{Karatzikos2010_PLB689-72}.
In the MDC-RMF models \cite{Lu2014_PRC89-014323}, the BCS approach are used to
treat the pairing correlations.
In the MDC-RHB models \cite{Zhao2016_in-prep}, the Bogoliubov transformation was implemented.
Both in MDC-RMF and in MDC-RHB models, we can use the $\delta$-force of zero-range or 
the separable pairing force of finite-range \cite{Tian2006_CPL23-3226,Tian2009_PLB676-44,Tian2009_PRC79-064301}.

To obtain a PES, i.e., $E = E( \{ \beta_{\lambda\mu} \})$, 
one can perform constraint calculations \cite{Ring1980}.
We have proposed a modified linear constraint method
in which the Routhian is calculated as,
\begin{equation}
 E^{\prime} = E_{{\rm RMF}} +
              \sum_{\lambda\mu} \frac{1}{2} C_{\lambda\mu}Q_{\lambda\mu} .
\end{equation}
In the $(n+1)$th iteration, the variable $C_{\lambda\mu}$ is determined by,
\begin{equation}
 C_{\lambda\mu}^{(n+1)} =
 C_{\lambda\mu}^{(n)} +
  k_{\lambda\mu} \left( \beta_{\lambda\mu}^{(n)} - \beta_{\lambda\mu} \right),
\end{equation}
where $\beta_{\lambda\mu}$ is the desired deformation,
$k_{\lambda\mu}$ is a constant and $C_{\lambda\mu}^{(n)}$ is the value in the $n$th iteration.

The RMF equations are solved iteratively. After a desired accuracy is
achieved, we can calculate various physical quantities.
For example, the total binding energy of the nucleus reads
\begin{eqnarray}
 E & = &
 \int d^{3}\bm{r}
  \left\{ \sum_{k}
           v_{k}^{2} \psi_{k}^{\dagger}
           \left( \bm{\alpha}\cdot\bm{p}+\beta M \right) \psi_{k} \right.
 \nonumber \\
 &  & \mbox{}
         +\frac{1}{2} \alpha_{ S} \rho_{ S}^{2} + \frac{1}{2} \alpha_{ V} \rho_{ V}^{2}
         +\frac{1}{2} \alpha_{TS} \rho_{TS}^{2} + \frac{1}{2} \alpha_{TV} \rho_{TV}^{2}
 \nonumber \\
 &  & \mbox{}
         +\frac{1}{3} \beta_{S} \rho_{S}^{3} + \frac{1}{4} \gamma_{S} \rho_{S}^{4}
                                             + \frac{1}{4} \gamma_{V} \rho_{V}^{4}
 \nonumber \\
 &  & \mbox{}
         +\frac{1}{2} \delta_{S} \rho_{S} \Delta\rho_{S}
         +\frac{1}{2} \delta_{V} \rho_{V} \Delta\rho_{V}
 \nonumber \\
 &  & \left. \mbox{}
         +\frac{1}{2} \delta_{TS} \rho_{TS} \Delta\rho_{TS}
         +\frac{1}{2} \delta_{TV} \rho_{TV} \Delta\rho_{TV}
         +\frac{1}{2}e\rho_{C}A
  \right\}
 \nonumber \\
 &  & \mbox{}
         + E_{{\rm pair}} + E_{{\rm c.m.}}
  ,
\end{eqnarray}
where $E_{{\rm pair}}$ is the pairing energy and
$E_{{\rm c.m.}}$ is the energy corresponding to the center of mass correction.
The intrinsic multipole moments are calculated from the density by
\begin{equation}
 Q_{\lambda\mu} = \int d^{3}\bm{r} \rho_{V}(\bm{r}) r^{\lambda} Y_{\lambda\mu}(\Omega),
\end{equation}
where $Y_{\lambda\mu}(\Omega)$ is the spherical harmonics.
The deformation parameter $\beta_{\lambda\mu}$ is obtained from
the corresponding multipole moment by
\begin{equation}
 \beta_{\lambda\mu} = \frac{4\pi} {3NR^{\lambda}} Q_{\lambda\mu},
\end{equation}
where $R= r_0 A^{{1}/{3}}$ is the radius of the nucleus, the
parameter $r_0=1.2$~fm
and $N$ represents proton, neutron or nucleon numbers.



\section{\label{sec:PES}PES and fission barriers of actinides}
\begin{figure*}
\begin{center}
\resizebox{0.85\textwidth}{!}{%
\includegraphics{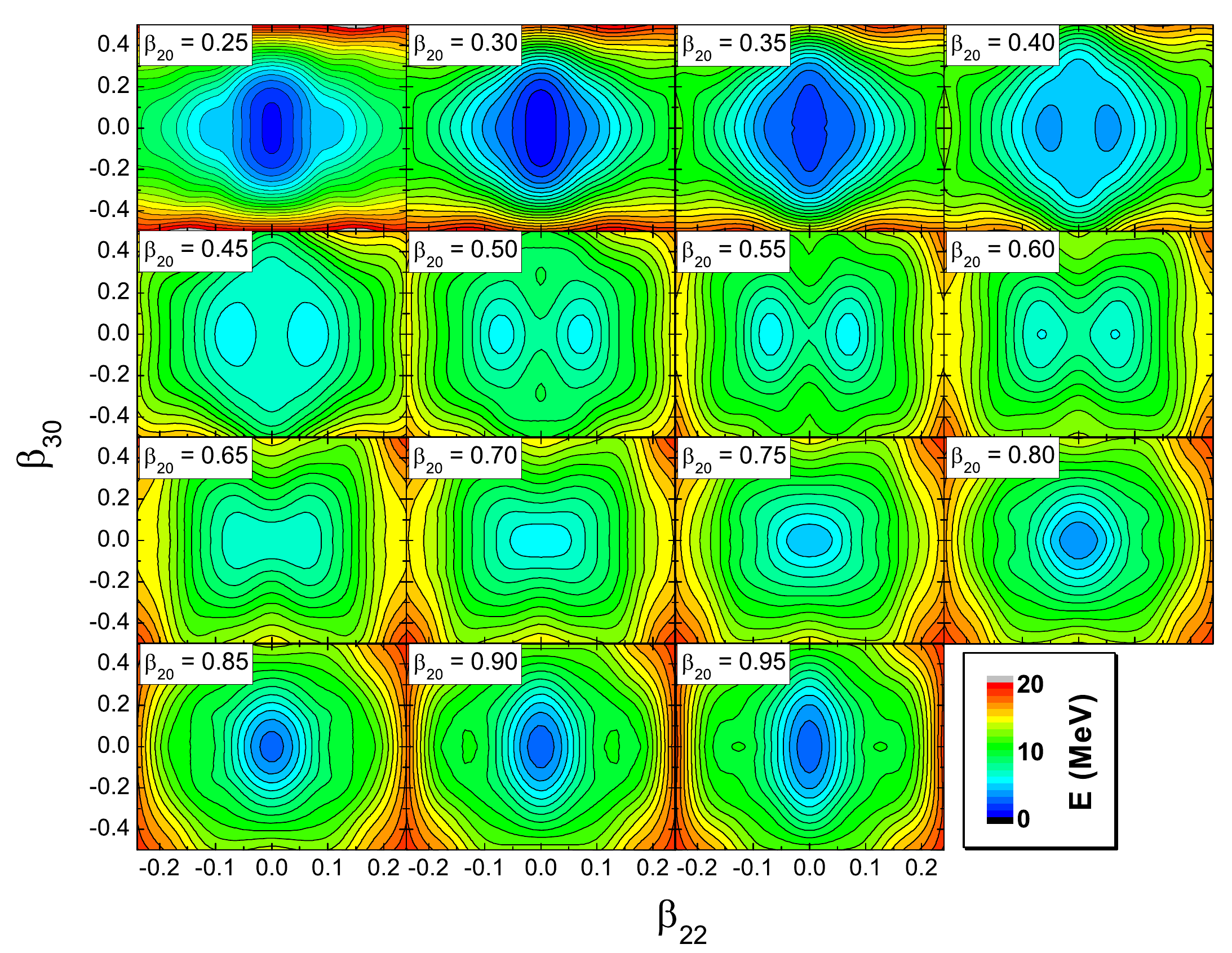}
}
\resizebox{0.85\textwidth}{!}{%
\includegraphics{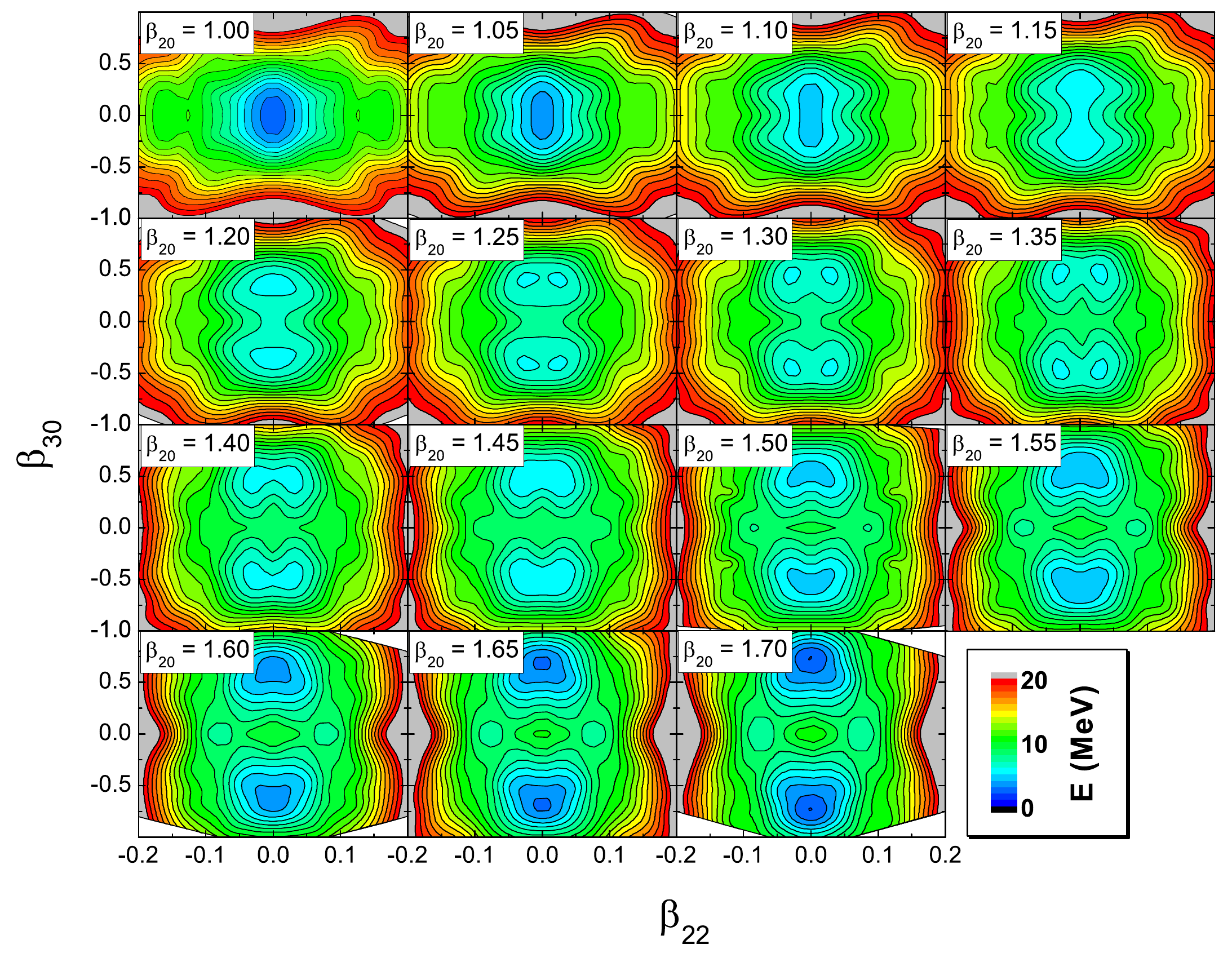}
}
\end{center}
\caption{(Color online)~\label{fig:d3plot}
Three-dimensional potential energy surface, i.e., $E=E(\beta_{20},\beta_{22},\beta_{30})$,
for $^{240}$Pu obtained from MDC-RMF calculations ($E$ is normalized
with respect to that of the ground state).
In each sub-figure, $\beta_{20}$ is fixed as indicated and 
the energy $E$ is displayed as a function of $\beta_{22}$ and $\beta_{30}$.
The contour interval is 1 MeV.
Taken from Ref.~\cite{Lu2014_PRC89-014323}.
}
\end{figure*}

In the study of nuclear fission, the potential energy surface (PES) 
is a crucial ingredient.
Physical observables, 
including the fragment mass distribution and the half-life, 
are determined to a large extent by the PES of a fissile nucleus.
Actinide nuclei, characterized by a double- or even triple-humped fission barriers, 
have been studied and used as the benchmark in many theoretical works
\footnote{Note that fission barriers can not be measured directly in experiment. 
One usually assumes a parabolic shape for the fission barrier and 
obtain the height and width of the barrier by fitting available experimental 
fission cross sections \cite{Capote2009_NDS110-3107}.
Therefore such barrier heights should be treated as ``empirical'' ones.}.
There have been very extensive studies on fission barriers and PES's for 
actinides as well as superheavy nuclei, see Ref.~\cite{Bender2003_RMP75-121}
for a review and Refs.~\cite{Abusara2010_PRC82-044303,%
Lu2012_PRC85-011301R,Lu2012_PhD,Lu2012_EPJWoC38-05003,Zhao2012_PRC86-057304,%
Abusara2012_PRC85-024314,Prassa2012_PRC86-024317,Afanasjev2012_IJMPE21-1250025,%
Afanasjev2013_ICFN5-303,Afanasjev2013_EPJWoC62-03003,%
Lu2014_PRC89-014323,Lu2014_PS89-054028,Lu2014_JPCS492-012014,%
Zhao2015_PRC91-014321} for recent progress.
In particular, in the framework of covariant density functional theory, 
Rutz et al. revealed the importance 
of the octupole shape on the second fission barrier in actinides \cite{Rutz1995_NPA590-680}
and Abusara et al. revealed the crucial role of triaxiality 
on the first fission barrier in actinides \cite{Abusara2010_PRC82-044303}.

We have obtained the PES's of actinide nuclei in 
three-dimensional deformation space, i.e., $E=E(\beta_{20}, \beta_{22}, \beta_{30})$,
with the MDC-RMF model \cite{Lu2012_PRC85-011301R,Lu2014_PRC89-014323}.
We found that for the second fission barriers (the outer one in many cases) in these actinides, 
the triaxial shape is very crucial.
Both the second barrier and the first one are lowered considerably due to the 
triaxial distortion as compared with results from axially symmetric calculations.
With the triaxial deformation considered, a better agreement 
with empirical values of the outer barrier heights in actinide nuclei was found.
By examining these PES's, one can learn a lot of interesting and useful
information concerning the shape and stiffness of the ground and 
isomeric states, the fission barrier heights and the lowering effect
of triaxial and octupole deformations on fission barriers.
Furthermore, by examining two-dimensional PES's and one-dimensional
potential energy curves (PECs) of some light actinides,
the third minima and third barriers in these surfaces (curves) were located 
and analyzed in detail \cite{Zhao2015_PRC91-014321}. 
We will present some of these results in this Section.

\subsection{\label{sec:240Pu}$^{240}$Pu: A typical example}

\subsubsection{Three-dimensional PES}

The MDC-RMF calculations were carried out for $^{240}$Pu in the following
deformation lattice:
\begin{itemize}
\item $\beta_{20} = 0.25$--1.70 with a step size of 0.05;
\item $\beta_{22} = 0   $--0.25 with a step size of 0.01;
\item $\beta_{30} = 0   $--0.50 with a step size of 0.05.
\end{itemize}
Totally 
26 (for $\beta_{22}$) $\times$ 
11 (for $\beta_{30}$) $\times$ 
30 (for $\beta_{20}$) = 8580 points were calculated.
The three-dimensional PES is shown in Fig.~\ref{fig:d3plot};
in each sub-figure the value of $\beta_{20}$ 
is fixed and the energy of $^{240}$Pu is shown as a function of
$\beta_{22}$ and $\beta_{30}$.
In these calculations, we used the effective interaction PC-PK1 
\cite{Zhao2010_PRC82-054319,Zhao2012_PRC86-064324}.
More details can be found in Refs.~\cite{Lu2012_PRC85-011301R,Lu2014_PRC89-014323}.

From the first three sub-figures in Fig.~\ref{fig:d3plot} with $\beta_{20}=0.25$, 0.30 and 0.35,
one finds that although the ground state with $\beta_{20}\approx 0.3$ is
a bit soft against the reflection asymmetric distortion,
its equilibrium shape is both reflection symmetric and axially symmetric.
When it is driven to be more elongated,
$^{240}$Pu becomes softer against the non-axial as well as octupole distortions.
From sub-figures with $\beta_{20}=0.40$--0.65, one finds that
there appears two symmetric minima with non-zero $\beta_{22}$. 
Although in these sub-figures all minima correspond to $\beta_{30}=0$, their softness 
against the reflection asymmetric distortion changes as $\beta_{20}$ increases.
The inner fission barrier $^{240}$Pu locates at $\beta_{20}\approx 0.60$.
In sub-figures with $\beta_{20}=0.70$--0.95,
the situation becomes much simpler, where the nucleus is again axially symmetric
and reflection symmetric.
Nevertheless, the trend to become softer against the octupole distortion 
can be seen from these sub-figures, 
which indicates that the reflection asymmetric shape becomes more relevant.

As $\beta_{20}$ becomes larger, there appears a superdeformed minimum 
which corresponds to the fission isomer of $^{240}$Pu; 
beyond the second minimum there exists the second saddle point in the potential energy surface.
From sub-figures with $\beta_{20} =$ 0.85--1.10, we see that,
$^{240}$Pu is reflection symmetric around the superdeformed minimum with 
$\beta_{20}\approx 0.95$,
but becomes softer against the $\beta_{30}$ distortion.
Starting from $\beta_{20}=1.15$, there exist two extrema with non-zero $\beta_{30}$.
Although these two extrema locate at the $\beta_{22}=0$ axis, i.e., 
the triaxial deformation is not important,
$^{240}$Pu becomes quite soft along the $\beta_{22}$ direction.
With $\beta_{20}$ further increasing ($1.25 \le \beta_{20} \le 1.35$), 
each minimum of reflection asymmetry splits into two minima with $\beta_{22} \ne 0$.
This is the lowering effect of triaxiality on the outer fission barrier in $^{240}$Pu.
Around this barrier, the energy is lowered by about 1 MeV by the triaxial distortion.
When $\beta_{20} \ge 1.6$ (the last three sub-figures in Fig.~\ref{fig:d3plot}), 
$^{240}$Pu becomes axially symmetric.

From these studies for $^{240}$Pu, we can draw the following conclusions
\cite{Lu2012_PRC85-011301R,Lu2014_PRC89-014323}:
(1) The shape of $^{240}$Pu in the ground state and in the superdeformed isomeric
state are axially symmetric and reflection symmetric;
(2) By examining the three-dimensional PES one can learn useful information on
the stability of $^{240}$Pu 
against the triaxial and octupole distortions:
Compared to the ground state, the fission isomeric state is much stiffer against 
both the non-axial and reflection asymmetric distortions; 
(3) While around the inner barrier it is triaxial and reflection symmetric,
$^{240}$Pu is both triaxial and reflection asymmetric around the second saddle point with
$\beta_{20} \approx 1.3$; 
(4) The triaxial deformation appears only around the fission barriers
and its effects are more pronounced around the first saddle point. 

\subsubsection{Two-dimensional PES}
 
\begin{figure}
\begin{center}
\resizebox{0.90\columnwidth}{!}{%
\includegraphics{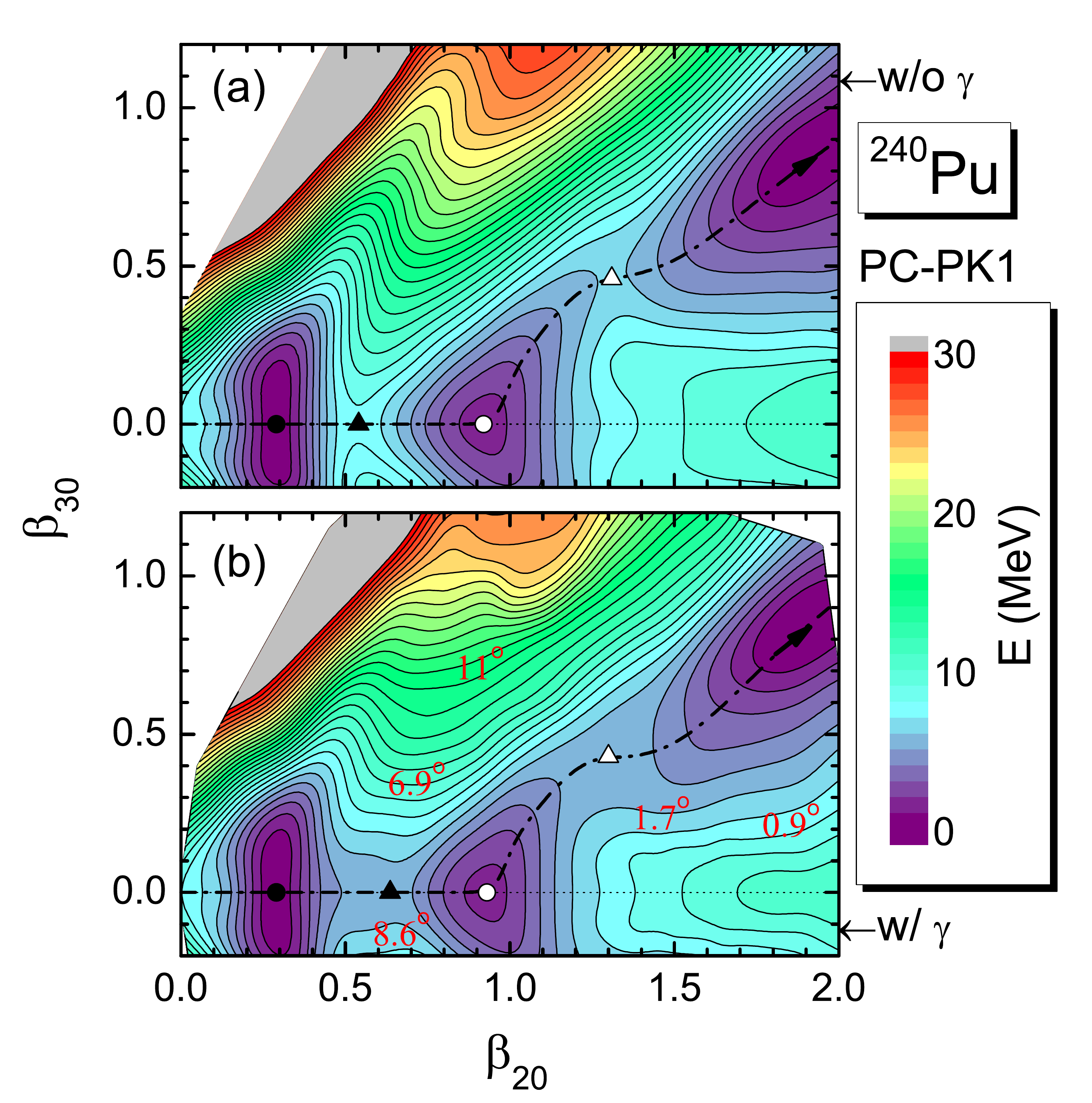}
}
\end{center}
\caption{\label{Pic:PU240_2d}(Color online) 
Two-dimensional potential energy surfaces of $^{240}$Pu [$E=E(\beta_{20},\beta_{30})$]
from MDC-RMF calculations without (a) and with (b) the triaxiality included
($E$ is normalized with respect to that of the ground state).
The values of $\gamma$ are shown in (b) for some points.
The dash-dotted line represents the static fission path. 
Full and open circles indicate the ground state and the fission isomer. 
Full and open triangles display the first and second saddle points. 
The contour interval is 1 MeV. 
Adapted from Ref.~\cite{Lu2012_PRC85-011301R}.
}
\end{figure}

The three-dimensional PES can be projected onto a two-dimensional deformation space
in several ways.
In Fig.~\ref{Pic:PU240_2d} we show two-dimensional PES [$E=E(\beta_{20},\beta_{30})$]
from MDC-RMF calculations.
When the triaxial distortion is allowed, the energy of $^{240}$Pu 
reaches the lowest value at each $(\beta_{20},\beta_{30})$ point during the iteration. 
At some points one gets non-zero $\beta_{22}$;
this means that the nucleus favors a triaxial shape. 
The triaxial shape mainly appears in two regions as seen in Fig.~\ref{Pic:PU240_2d}(b). 
The first region starts from the first saddle point with $\beta_{20}\approx 0.60$ and 
continues almost vertically up to $\beta_{30} \approx 1.0$: 
The values of $\beta_{22}$ are around 0.06--0.12
and correspondingly, $\gamma \approx$ 7--11$^{\circ}$;  
the triaxial distortion lowers the binding energy by 2 MeV. 
The second region is around the outer fission barrier: 
The $\beta_{22} \approx 0.02$--0.03 corresponding to $\gamma \approx$ 1--2$^{\circ}$. 
At the second saddle point, the binding energy is lowered by around 1 MeV  
due to the triaxial shape. 

\begin{figure}[h]
\begin{center}
\resizebox{0.90\columnwidth}{!}{%
 \includegraphics{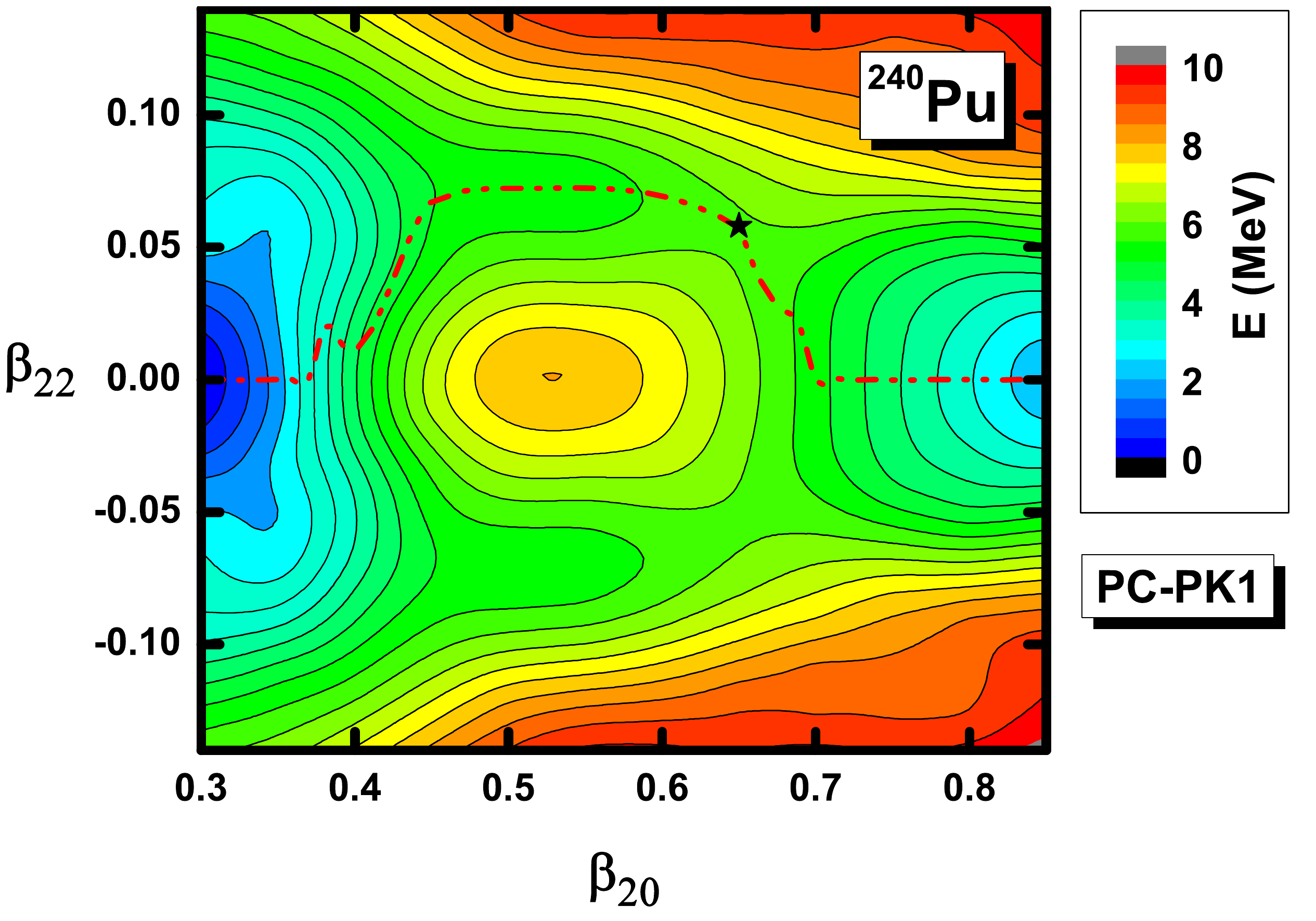} }
\end{center}
\caption{\label{Pic:PU240_2dI}(Color online)
Two-dimensional potential energy surface of $^{240}$Pu [$E=E(\beta_{20},\beta_{22})$]
around the inner barrier
($E$ is normalized with respect to that of the ground state).
The dash-dotted line represents the static fission path. 
The full star denotes the first saddle point.
The contour interval is 0.5 MeV.
Taken from Ref.~\cite{Lu2014_PS89-054028}.
}
\end{figure}

\begin{figure}[h]
\begin{center}
\resizebox{0.90\columnwidth}{!}{%
 \includegraphics{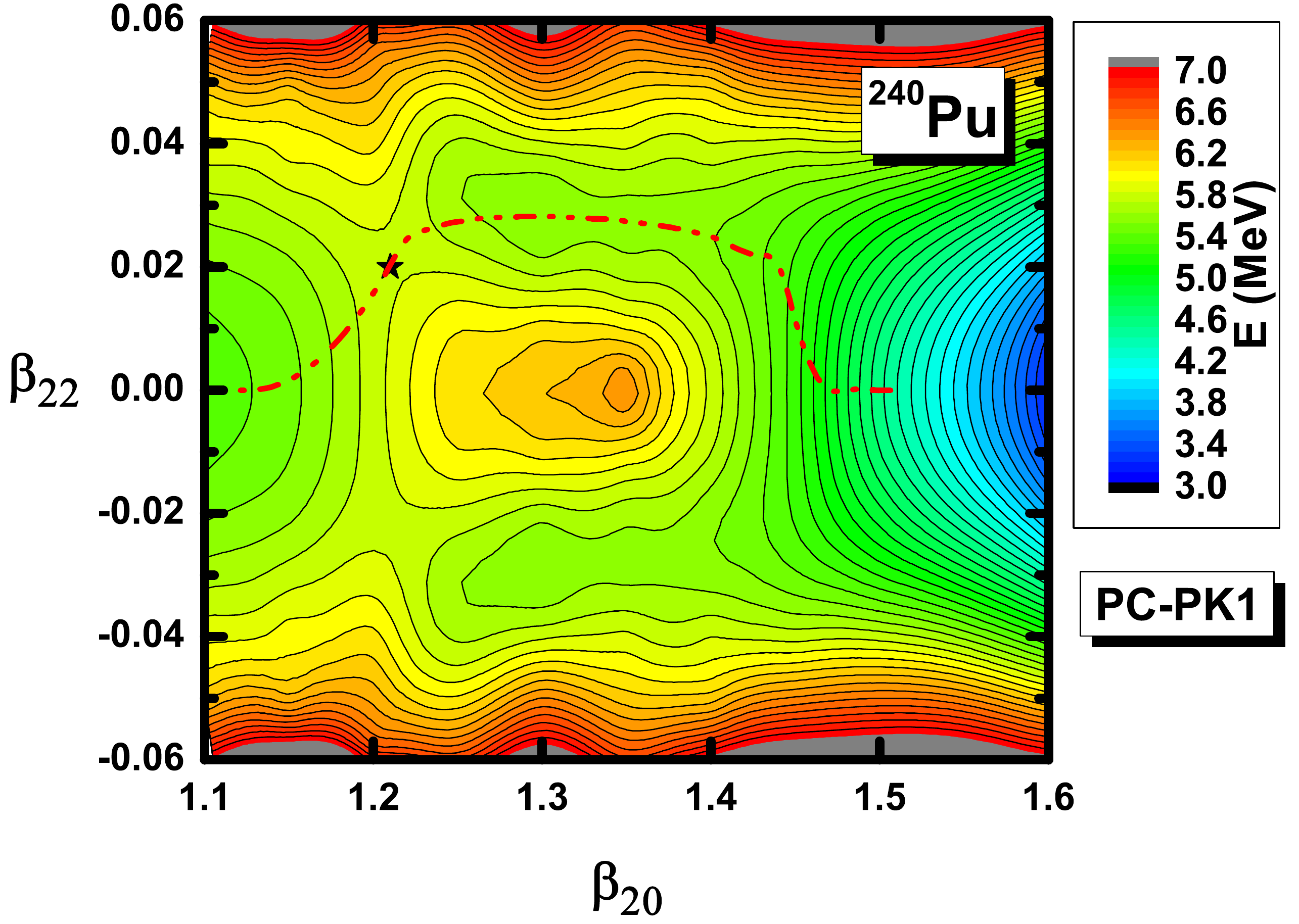} }
\end{center}
\caption{\label{Pic:PU240_2dO}(Color online)
Two-dimensional potential energy surface of $^{240}$Pu [$E=E(\beta_{20},\beta_{22})$]
around the second fission barrier
($E$ is normalized with respect to that of the ground state).
The dash-dotted line represents the static fission path. 
The full star denotes the second saddle point.
The contour interval is 0.1 MeV.
Adapted from Ref.~\cite{Lu2014_PS89-054028}.
}
\end{figure}

Next we focus on the PES's [$E=E(\beta_{20},\beta_{22})$] around the two
saddle points
which are shown in Figs.~\ref{Pic:PU240_2dI} and~\ref{Pic:PU240_2dO}.
Starting from its ground state which is axially symmetric, 
the nucleus $^{240}$Pu evolves to the isomeric state through the triaxial valley. 
The inner barrier is located at $\beta_{20} \approx 0.65$ and $\beta_{22} \approx 0.06$. 
The isomeric state keeps an axially symmetric shape.
With $\beta_{20}$ further increasing, the nucleus enters again a valley with triaxiality, 
then arrives at fission configurations. 
The second fission barrier locates at 
$(\beta_{20},\beta_{22},\beta_{30}) \approx (1.21, 0.02, 0.37)$.

\subsubsection{One-dimensional PEC}

\begin{figure}
\begin{center}
\resizebox{0.9\columnwidth}{!}{%
 \includegraphics{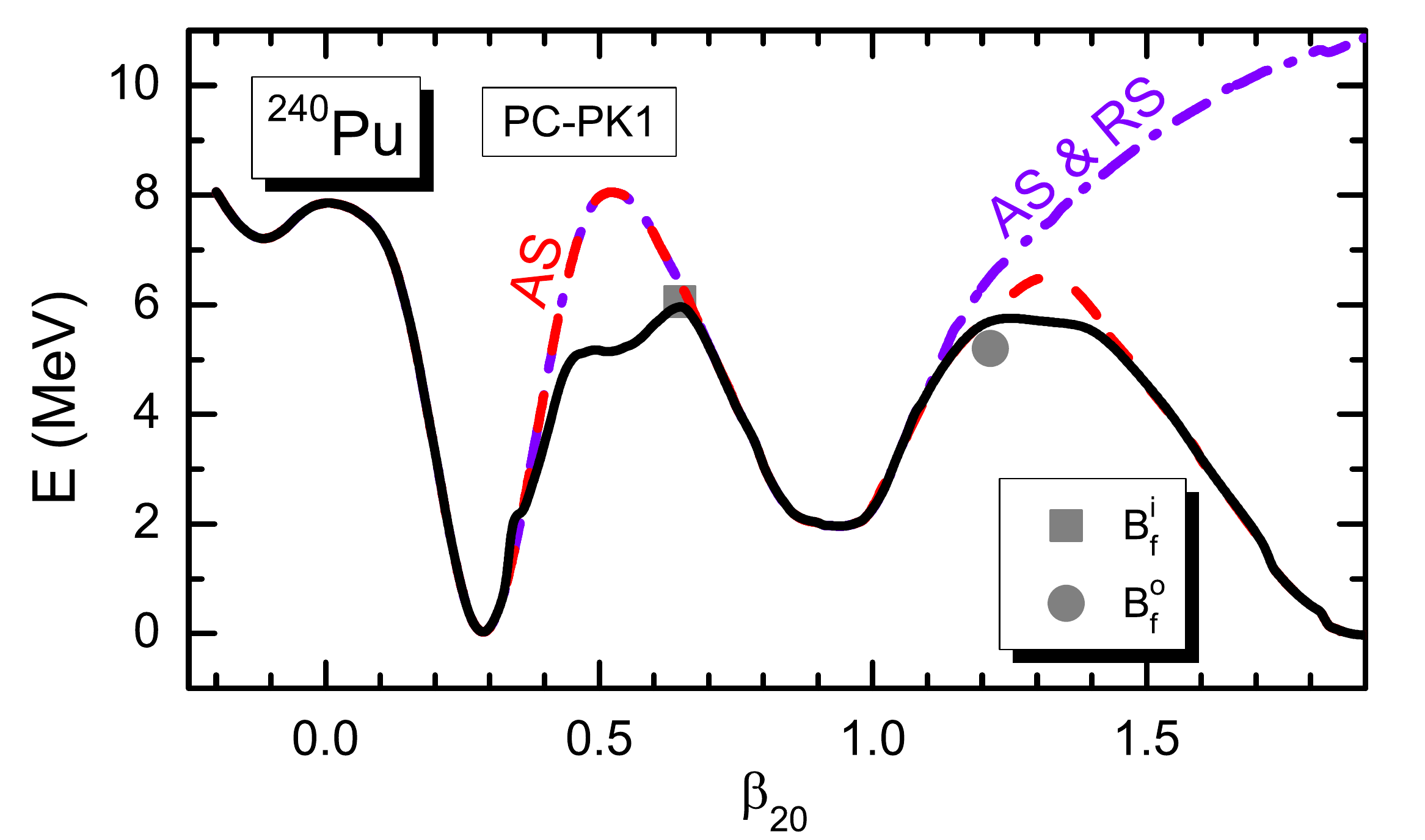} }
\end{center}
\caption{\label{Pic:PU240-1d}(Color online)
Potential energy curve of $^{240}$Pu [$E=E(\beta_{20})$] obtained from MDC-RMF calculations 
with various symmetries imposed
($E$ is normalized with respect to that of the ground state): 
(1) The solid black curve [triaxial (TA) and reflection asymmetric (RA)]; 
(2) The red dashed curve [axial symmetric (AS) but RA];
(3) The violet dot-dashed line [AS \& reflection symmetric (RS)].
The grey square (circle) denotes the empirical inner (outer) barrier height.
Adapted from Ref.~\cite{Lu2012_PRC85-011301R}.
}
\end{figure}

One can further project the three-dimensional or two-dimensional PES 
onto one shape degree of freedom and obtain one-dimensional potential energy curves (PECs).
In Fig.~\ref{Pic:PU240-1d} we show the PEC 
from $\beta_{20} \approx -0.2$ to the fission configuration 
calculated with various symmetries imposed
for $^{240}$Pu: the axial symmetry (AS) or triaxial (TA) 
symmetry together with reflection symmetry (RS) or reflection asymmetry (RA). 

In Fig.~\ref{Pic:PU240-1d} one can clearly see the importance of the triaxiality
around the inner fission barrier as well as the crucial role of the axial octupole $Y_{30}$ 
shape around the outer fission barrier: 
The triaxial deformation reduces the height of the inner fission barrier considerably 
(more than 2 MeV), thus leading to a good agreement with the empirical value;
the reflection asymmetric shape becomes very important beyond the second minimum and 
lowers the second fission barrier. 

In addition, we have revealed in Ref.~\cite{Lu2012_PRC85-011301R} 
for the first time that the second fission
barrier is also lowered by the triaxiality and around 1 MeV is gained
in energy for the second saddle point when the triaxial distortion is allowed. 
This results in a better reproduction of the empirical height for
the outer fission barrier.
Needless to say, this feature can be revealed only when the axial and 
reflection symmetries are broken simultaneously.

\subsection{\label{sec:PES-double}%
Double-humped PES's and fission barriers of even-even actinides} 

\subsubsection{One-dimensional PECs around the barriers}

\begin{figure*}
\begin{center}
\resizebox{0.85\textwidth}{!}{%
\includegraphics{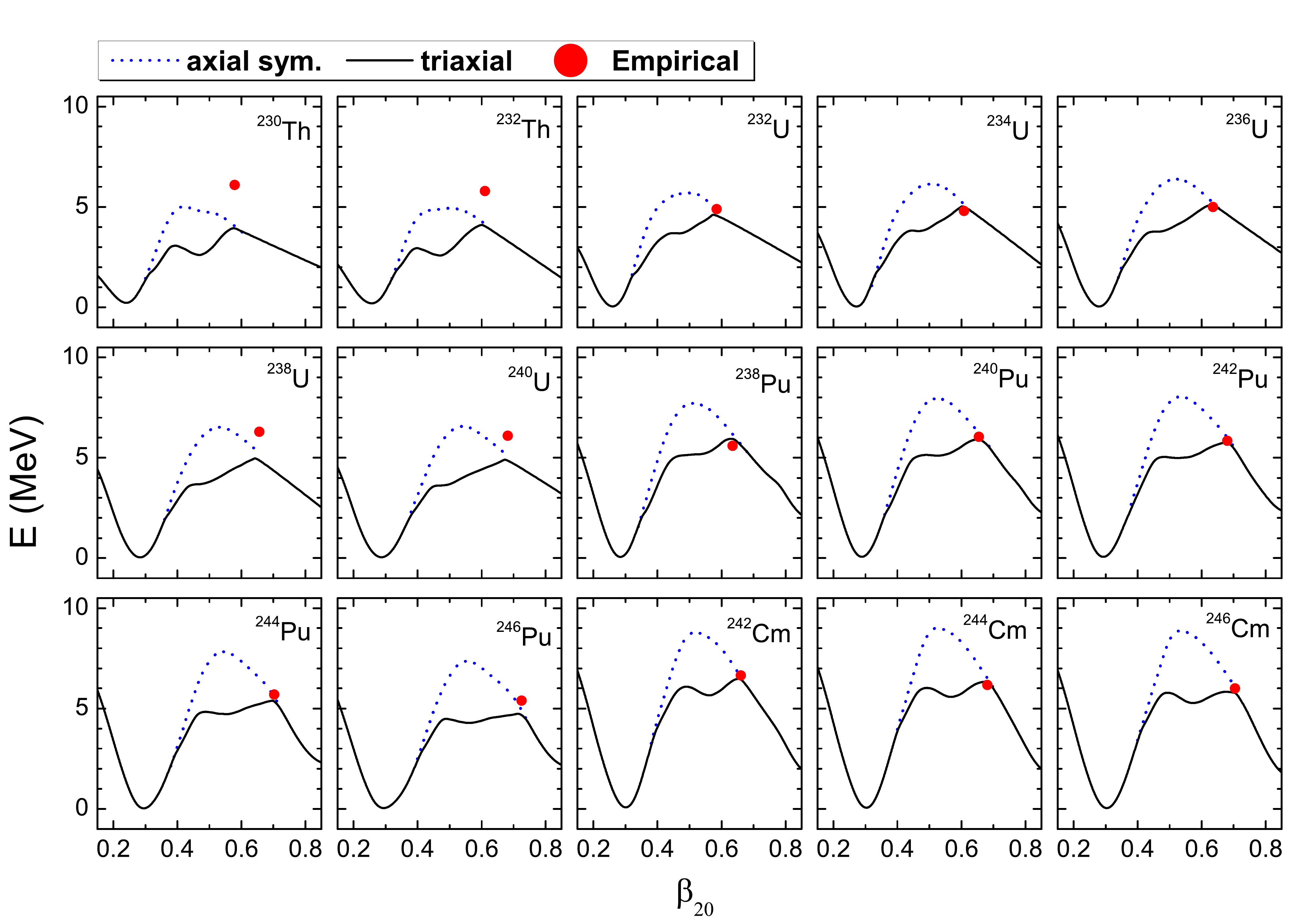}
}
\end{center}
\begin{center}
\resizebox{0.85\textwidth}{!}{%
\includegraphics{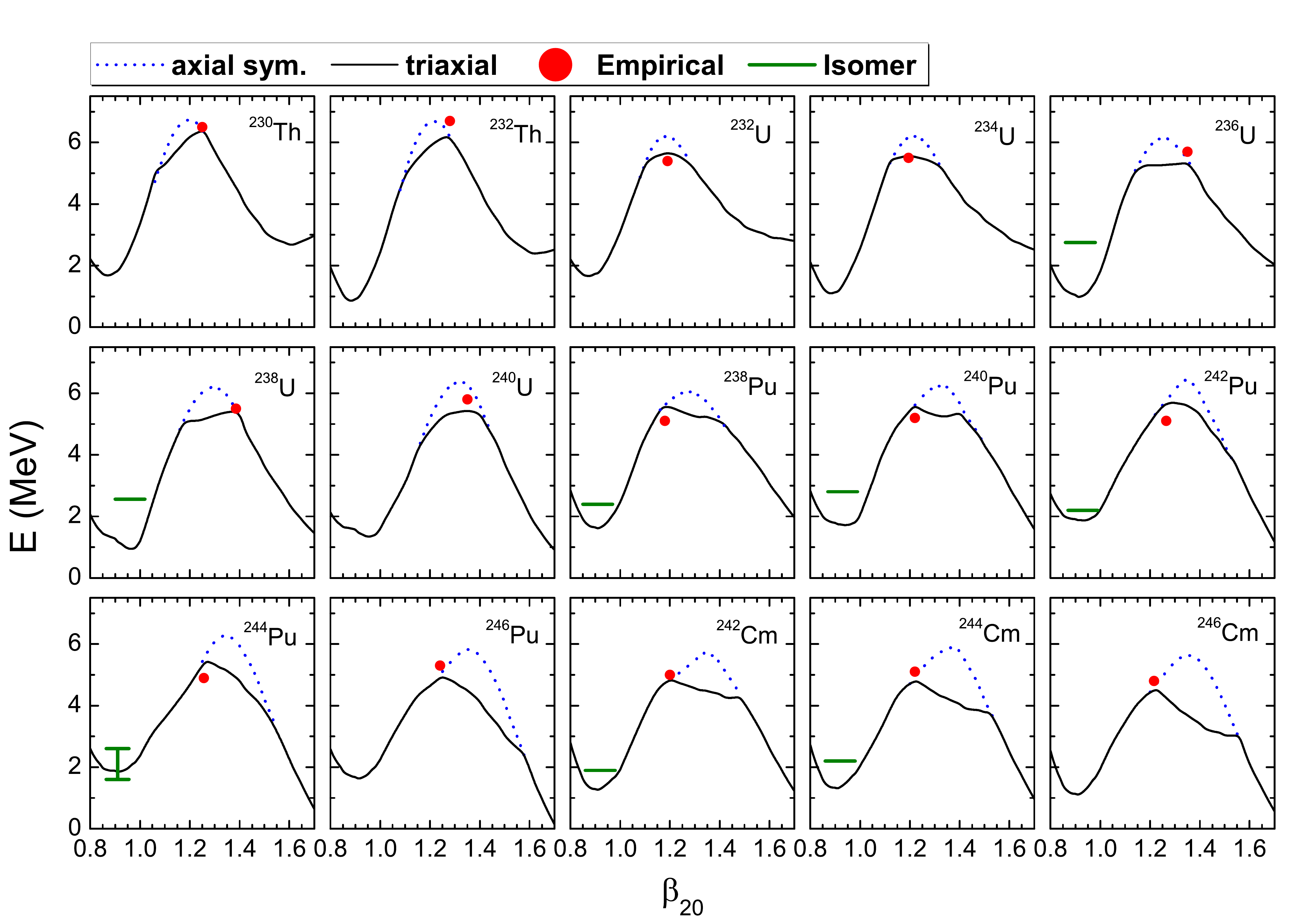}
}
\end{center}
\caption{(Color online)~\label{fig:actinidePESinner}
Potential energy curves [$E=E(\beta_{20})$] obtained from MDC-RMF 
calculations for even-even actinide nuclei
($E$ is normalized with respect to that of the ground state).
The upper panel shows PECs around the ground states and the first fission barriers and
the lower panel shows PECs around the fission isomers and the second fission barriers.
Dotted curves represent the axially symmetric results and
solid curves display results from the triaxial calculations.
The empirical values for fission barrier heights are taken from 
Ref.~\cite{Capote2009_NDS110-3107} and shown by red dots.
The green horizontal lines show experimental energies for fission isomers 
\cite{Singh2002_NDS97-241}.
Taken from Ref.~\cite{Lu2014_PRC89-014323}.
}
\end{figure*}

Constraint calculations for a three-dimensional PES are very time-consuming.
From the three-dimensional calculations of $^{240}$Pu, 
we can obtain useful experiences on the roles of various nuclear shapes
for actinide nuclei.
Guided by the features found in the one-, two- and three-dimensional 
PES's of $^{240}$Pu,
potential energy curves and fission barriers of even-even actinide nuclei 
have been systematically 
studied in Ref.~\cite{Lu2014_PRC89-014323}.

An actinide nucleus has a reflection symmetric but triaxial shape 
around the first fission barrier. 
Therefore one can make a one-dimensional constraint calculation with
the triaxial deformation considered but not the octupole shape.
In the upper panel of Fig.~\ref{fig:actinidePESinner} we present
PECs covering the ground state and the inner fission barrier for 
some even-even actinide nuclei.
Dotted curves represent the axially symmetric results and
solid curves display results from the triaxial MDC-RMF calculations.
The empirical fission barrier heights \cite{Capote2009_NDS110-3107}
are shown for comparison.
It can be clearly seen that the triaxial distortion lowers the first fission barrier
of these even-even actinides by about 1 to 4 MeV.
The agreement of the calculated fission barrier heights with the empirical ones
is good for most of the nuclei studied here but $^{230,232}$Th and $^{238,240}$U. 
Possible reasons for these discrepancies will be discussed later.

It is more complicated to determine the second fission barrier height
because more shape degrees of freedom become
relevant and there are often two or more static fission paths.
What we have done was the following~\cite{Lu2012_PRC85-011301R,Lu2014_PRC89-014323}:
(1) The axial symmetry is assumed and a two-dimensional constraint calculation
in the $(\beta_{20},\beta_{30})$ plane is made;
(2) From the two-dimensional calculation the lowest static fission path
$\beta_{30}^\mathrm{lowest}(\beta_{20})$ is approximately identified;
(3) Along this fission path, we perform MDC-RMF calculations 
with $\beta_{20}$ constrained to a given value and
with both octupole and triaxial deformations allowed. 
$\beta_{30}^\mathrm{ini.} = \beta_{30}^\mathrm{lowest}(\beta_{20})$
and a small $\beta_{22}$ are taken as
the initial deformation parameters;
(4) In this one-dimensional PEC, the second saddle point
is located and the second barrier height is extracted.

PECs in the region of the isomeric state and the outer fission barrier
for even-even actinide nuclei 
are shown in the lower panel of Fig.~\ref{fig:actinidePESinner}.
In addition, experimental values of the energies for fission isomers \cite{Singh2002_NDS97-241}
are also shown by short horizontal lines.
One finds that
the triaxiality can lower the second fission barrier by about 0.5 to 1 MeV
for most of the actinide nuclei investigated here. 
The fission barrier heights obtained in MDC-RMF calculations 
with the triaxial deformation included 
are in good agreements with the empirical values for these even-even actinides.

\subsubsection{Heights of the first and second fission barriers}

\begin{figure}
\begin{center}
\resizebox{0.8\columnwidth}{!}{%
\includegraphics{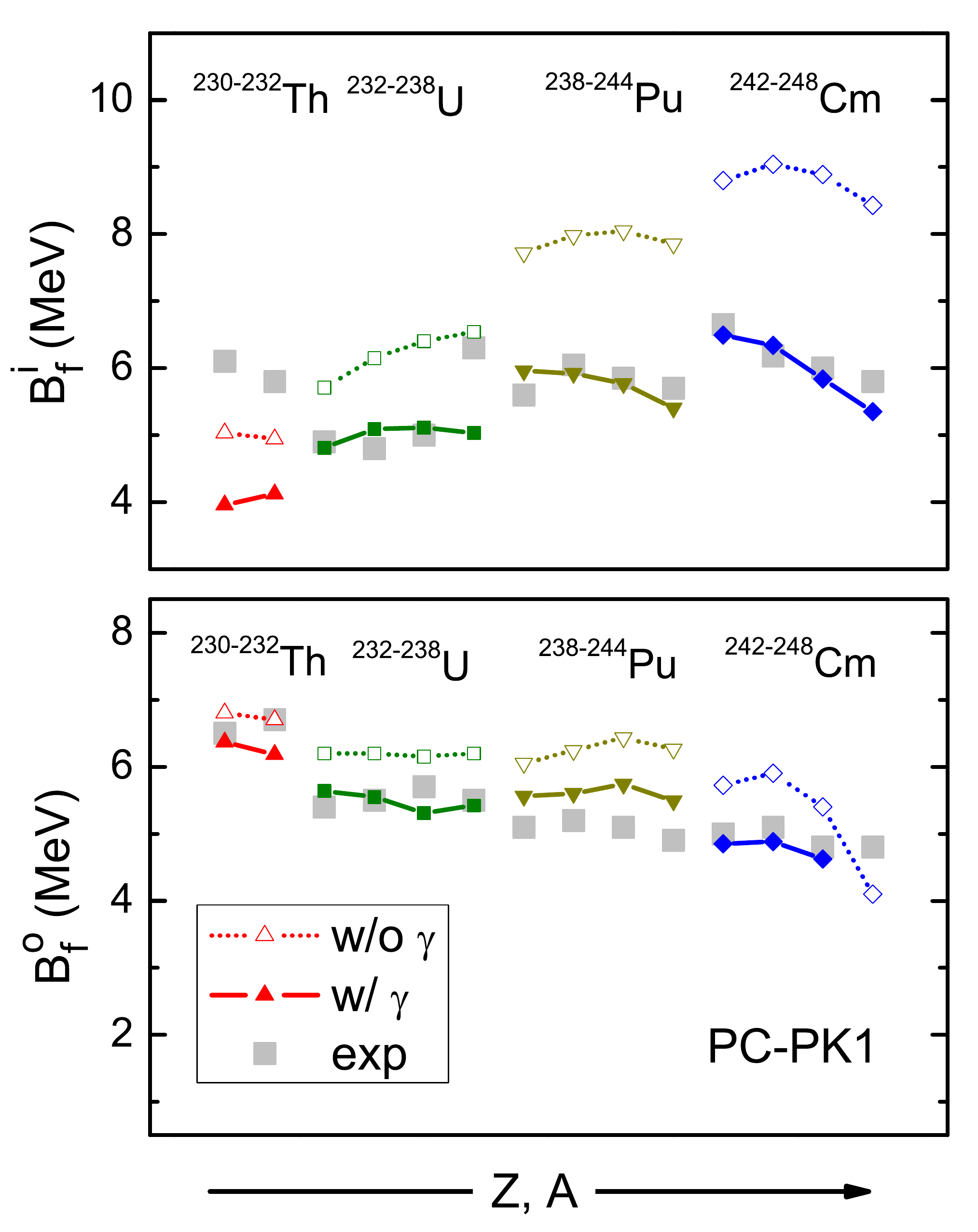}
}
\end{center}
\caption{\label{fig:The-calculated-inner}(Color online) 
The heights of inner ($B^\mathrm{i}_\mathrm{f}$, the upper panel) 
and outer ($B^\mathrm{o}_\mathrm{f}$, the lower panel) 
fission barriers for some even-even actinide nuclei. 
Open (full) symbols denotes results without (with) the triaxial distortion allowed.
The empirical barrier heights are taken from Ref.~\protect\cite{Capote2009_NDS110-3107}
and shown by grey squares.
Taken from Ref.~\cite{Lu2012_PRC85-011301R}.
}
\end{figure}

The heights for inner fission barriers $B^\mathrm{i}_\mathrm{f}$ calculated 
with the MDC-RMF models
are given in Fig.~\ref{fig:The-calculated-inner} (the upper panel).
The corresponding empirical values
are taken from Ref.~\cite{Capote2009_NDS110-3107} and shown for comparison.
In this figure one finds that the triaxial distortion lowers systematically
the inner barrier heights by 1--4 MeV in these actinide nuclei.
Generally speaking, the agreement between the calculated barrier heights
and the empirical ones is fairly good; but this is not the case
for $^{230,232}$Th and $^{238}$U. 
The inner barrier heights calculated from our model are smaller 
than the empirical values by around 1 MeV for $^{230,232}$Th
if the triaxial deformation is not considered. 
When the triaxial deformation is included, 
the calculated heights of inner barriers are smaller 
than the empirical values by around 2 MeV for these two nuclei. 
Comparing the two panels in Fig.~\ref{fig:The-calculated-inner}, one can
find that 
the inner barrier is lower than the outer one for $^{230,232}$Th.
That is, the inner fission barrier is not the primary one and
this may bring in some uncertainties when 
the parameters of inner fission barrier are determined \cite{Samyn2005_PRC72-044316}. 
Note that the Skyrme-Hartree-Fock-Bogoliubov calculations gave similar results 
for $^{230,232}$Th \cite{Samyn2005_PRC72-044316}
and another calculation with the covariant density functional theory also 
gave a very low inner barrier for $^{232}$Th \cite{Abusara2010_PRC82-044303}. 
For $^{238}$U, the inner fission barrier height calculated from 
the axial calculation is already quite comparable with the empirical height. 
The triaxial deformation further lowers this barrier by 1.5 MeV
and leads to a disagreement with the empirical value.  

In Fig.~\ref{fig:The-calculated-inner} we also show the
heights of outer fission barriers $B^\mathrm{o}_\mathrm{f}$ (the lower panel) 
and make comparison with empirical heights. 
For most of the nuclei we focused on, the triaxial distortion lowers the outer 
fission barrier by 0.5 to 1 MeV which is about 10 to 20\% of the barrier height.
With the triaxiality considered, the calculation agrees better with 
the empirical values for most of even-even actinides and the only exception 
is $^{248}$Cm. 
The reason for this discrepancy has been attributed to
the competition between two or more static fission paths beyond the first barrier 
\cite{Lu2012_PRC85-011301R}. 

We have also examined the dependence of the above conclusions on
the effective interactions used in MDC-RMF calculations. 
It was found that the lowering effect of the triaxial shape on the outer
fission barrier is still there when effective interactions other than PC-PK1 
were used.

\subsection{\label{sec:3rd-barrier}The third minima and barriers in light actinide nuclei}

As discussed in Sections~\ref{sec:240Pu} and \ref{sec:PES-double},
the potential energy surface of actinide nuclei shows a double-humped feature
along the fission path.
In as early as 1970s, by using macroscopic-microscopic models, M\"oller et al.
have predicted that a third minimum may exist beyond the second barrier in 
the PES's of some actinide nuclei
 \cite{Moeller1972_NPA192-529,%
Moeller1973_IAEA-SM-174-202,Moeller1974_NPA229-269}. 
The existence of such third minima has been used to account for the thorium anomaly 
\cite{Moeller1973_IAEA-SM-174-202,%
Moeller1974_NPA229-269,Bhandari1989_PRC39-917}.
High resolution fission cross section measurements were later
performed for $^{237}$U and $^{230-233}$Th and these measurements
indicate that there exists shallow third minima in these nuclei 
\cite{Blons1975_PRL35-1749,Blons1978_PRL41-1282,%
Blons1988_NPA477-231,Knowles1982_PLB116-315,Findlay1986_NPA458-217,%
Zhang1986_PRC34-1397,Yoneama1996_NPA604-263,Sin2006_PRC74-014608}.

In Ref.~\cite{Bengtsson1987_NPA473-77}, the PES's of Ra to Th isotopes were 
studied by adopting a macroscopic-microscopic model in which
a modified harmonic oscillator potential was used.
A third minimum was predicted for many nuclei at very large quadrupole deformation
and
the third potential pocket could be as deep as 1.5 MeV.
Later \'{C}wiok et al. investigated the potential energy surfaces of
some even-even actinides and found that
quite deep minima may exist in many of these nuclei
\cite{Cwiok1994_PLB322-304}.

In Refs.~\cite{Krasznahorkay1998_PRL80-2073,Krasznahorkay1999_PLB461-15,%
Csatlos2005_PLB615-175,Csige2009_PRC80-011301R,Csige2013_PRC87-044321},
experimental evidences for the third minima and hyperdeformed states 
in U isotopes were reported.
The excitation energies of the third saddle point and the third minimum in
these nuclei were deduced as 
$B_{\rm III} \approx 6$ MeV and $E_{\rm III}=3$--4 MeV, respectively.
This corresponds to third pockets with a depth around 2--3 MeV.
These results support predictions made in Ref.~\cite{Cwiok1994_PLB322-304},
but are consistent neither with the experiments concerning $^{230-233}$Th and
$^{237}$U \cite{Blons1975_PRL35-1749,Blons1978_PRL41-1282,%
Blons1988_NPA477-231,Knowles1982_PLB116-315,Findlay1986_NPA458-217,%
Zhang1986_PRC34-1397,Yoneama1996_NPA604-263}
nor with the theoretical results given in 
Refs.~\cite{Moeller1972_NPA192-529,%
Moeller1973_IAEA-SM-174-202,Moeller1974_NPA229-269,Bengtsson1987_NPA473-77}.

The macroscopic-microscopic model with more shape degrees of freedom included
was used to study PES's of actinide nuclei 
\cite{Kowal2012_PRC85-061302R,Jachimowicz2013_PRC87-044308}.
In particular, the authors of Refs.~\cite{Kowal2012_PRC85-061302R,Jachimowicz2013_PRC87-044308}
included the $\beta_{10}$ deformation which lowers the third barrier substantially
and leads to the disappearance of the third pockets in many actinide nuclei.
Third minima, which are quite shallow and the depth are less than 400 keV, 
were predicted only in $^{230,232}$Th \cite{Kowal2012_PRC85-061302R,Jachimowicz2013_PRC87-044308}.
Furthermore, Ichikawa et al. carried out 
a systematic study of PES's of actinide nuclei by using the finite-range liquid-drop model 
\cite{Moeller2009_PRC79-064304} and
found that the third pockets for light Th and U isotopes 
are also quite shallow and the depths are less than 1 MeV \cite{Ichikawa2013_PRC87-054326}.

Besides various versions of macroscopic-microscopic models, 
self-consistent mean field approaches,
e.g., 
Hartree-Fock or Hartree-Fock-Bogoliubov models 
with Skyrme forces 
\cite{Bonneau2004_EPJA21-391,Samyn2005_PRC72-044316,McDonnell2013_PRC87-054327}
and the Gogny force 
\cite{Berger1989_NPA502-85,Delaroche2006_NPA771-103,Dubray2008_PRC77-014310} 
and
the relativistic mean field model \cite{Rutz1995_NPA590-680},
have also been adopted to investigate PES's of actinide nuclei.
Generally speaking, no very deep third minima were predicted in
such studies.

The MDC-RMF model has been used to examine the occurrence and features of 
the third minima and the third barriers in the PES's of some light actinides
in Ref.~\cite{Zhao2015_PRC91-014321}.
We will present these results in this Section.

\begin{figure*}
\begin{center}
\resizebox{0.85\textwidth}{!}{%
 \includegraphics{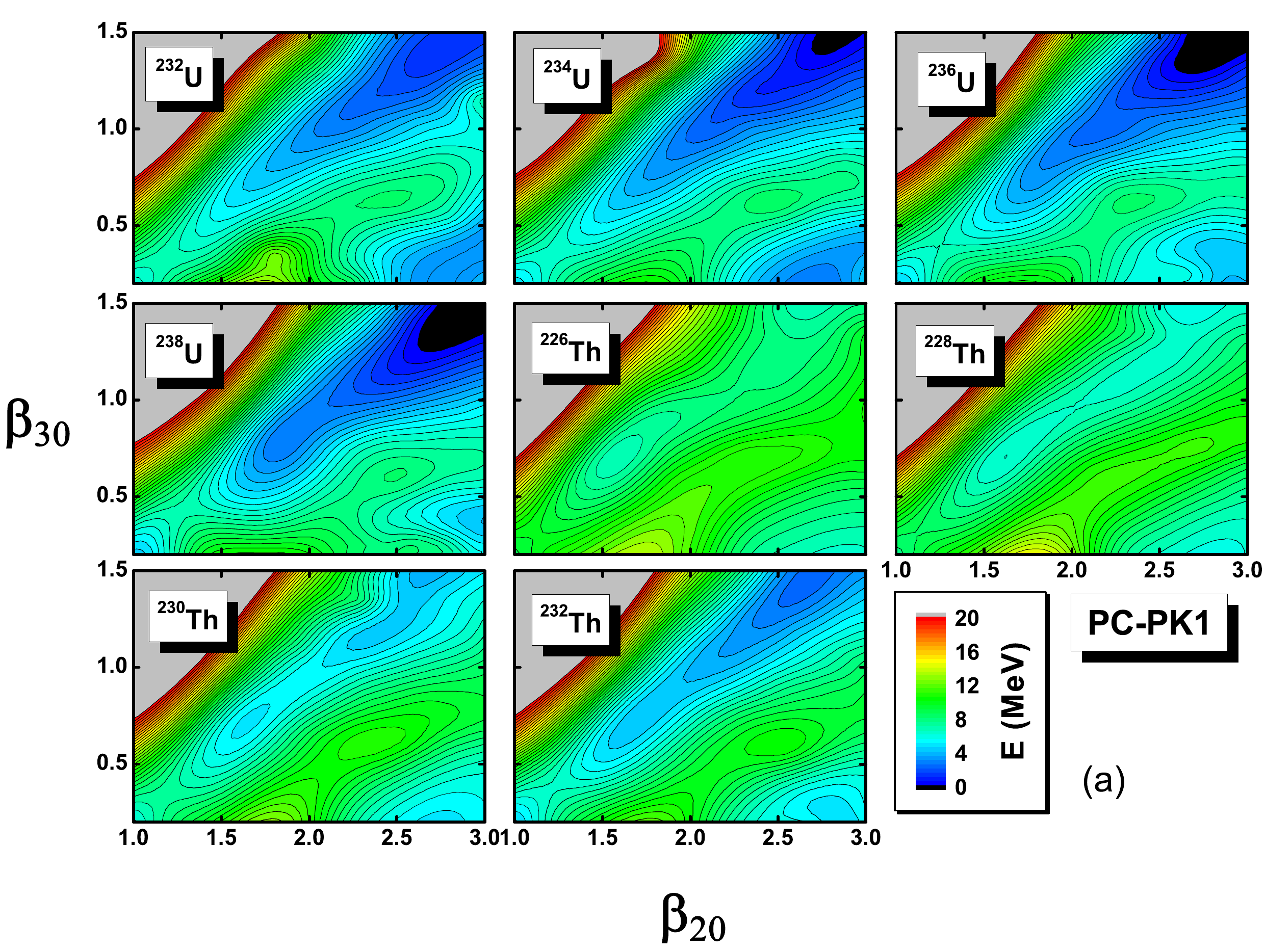}
}
\end{center}
\begin{center}
\resizebox{0.85\textwidth}{!}{%
 \includegraphics{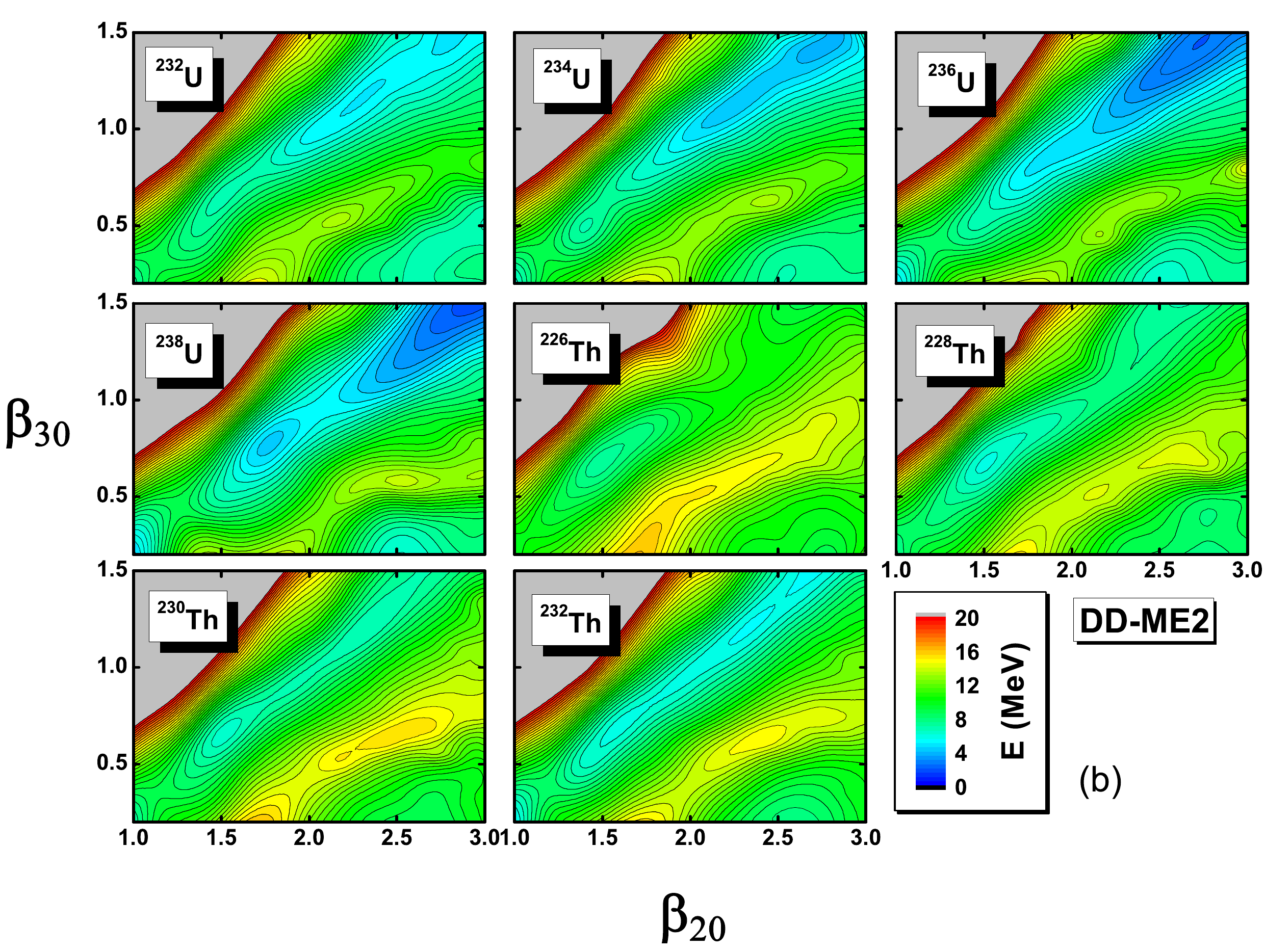}
}
\end{center}
\caption{(Color online)~\label{fig:pes}%
Potential energy surfaces of U and Th isotopes [$E=E(\beta_{20},\beta_{30})$] 
calculated by using the MDC-RMF models
with the relativistic functionals PC-PK1 (a) and DD-ME2 (b) 
($E$ is normalized with respect to that of the ground state). 
The contour interval is 0.5 MeV.
Taken from Ref.~\cite{Zhao2015_PRC91-014321}.
}
\end{figure*}

\subsubsection{Two-dimensional PES's beyond the second fission barrier}

We investigated even-even Th and U isotopes, namely, 
$^{226,228,230,232}$Th
and
$^{232,234,236,238}$U.
In Fig.~\ref{fig:pes} the PES's in the $(\beta_{20},\beta_{30})$ plane 
($\beta_{20} \approx 1.0$--3.0 and $\beta_{30} \approx 0.3$--1.5) are displayed
for these nuclei.
These PES's are obtained by using effective interactions PC-PK1 and DD-ME2 
\cite{Lalazissis2005_PRC71-024312}, respectively.
From Fig.~\ref{fig:pes}, we can locate the second barrier and if they exist,
the third barrier and the third minimum.

For $^{232,234,236,238}$U, the two functionals PC-PK1 and DD-ME2 give different
predictions concerning the existence of third minima.
There are no noticeable third minimum in the PES's from PC-PK1, 
as is seen in Fig.~\ref{fig:pes}(a).
The second saddle points of $^{232,234,236,238}$U locate around
$\beta_{20}\approx$ 1.2--1.3 and $\beta_{30}\approx$ 0.3--0.4.
With $\beta_{20}$ and $\beta_{30}$ further increasing, the energy becomes
smaller and smaller along the lowest fission path.
However, the PES's ($\beta_{20} > 1.5$ and $\beta_{30} > 0.5$) calculated 
with DD-ME2 are different from those with PC-PK1, as shown in Fig.~\ref{fig:pes}(b).
With the exception of $^{236}$U, one finds a third minimum in the PES's of 
other U isotopes, i.e., $^{232,234,238}$U.
The third pocket is quite deep for $^{238}$U, but
very shallow for $^{232}$U and $^{234}$U.
Similar results have been obtained by using 
the macroscopic-microscopic model 
\cite{Kowal2012_PRC85-061302R,Jachimowicz2013_PRC87-044308,%
Ichikawa2012_PRC86-024610,Ichikawa2013_PRC87-054326}
and the Skyrme Hartree-Fock-Bogoliubov model \cite{McDonnell2013_PRC87-054327}.

Compared to U isotopes, more pronounced third minima were predicted in 
Th isotopes with both PC-PK1 and DD-ME2 functionals.
For $^{226}$Th, the depth of the third pocket could be 1--2 MeV.
With $N$ increasing, the height of the third barrier and the energy of 
the third minimum both decrease and the difference between them becomes smaller,
i.e., the third well becomes shallower.
These tendencies have also been discussed in Ref.~\cite{Ichikawa2013_PRC87-054326}.
For $^{230}$Th, there is a shallow third well with a depth less than 1 MeV.
For $^{232}$Th, the third minimum disappears completely in the PES calculated
from the PC-PK1 functional 
while there is a third well, though quite shallow, in the PES obtained from the effective
interaction DD-ME2.

From Fig.~\ref{fig:pes} we have concluded that  more pronounced third minima
are predicted with the functional DD-ME2 than PC-PK1. 
In addition to the lowest fission path, 
one can also notice other possible paths along which there
appear several saddle points and shallow minima.
These fine structures in the PES's of actinides deserve more detailed
investigations.

\subsubsection{One-dimensional PECs around the third minimum and the third barrier}

\begin{figure*}
\begin{center}
\resizebox{0.85\textwidth}{!}{%
 \includegraphics{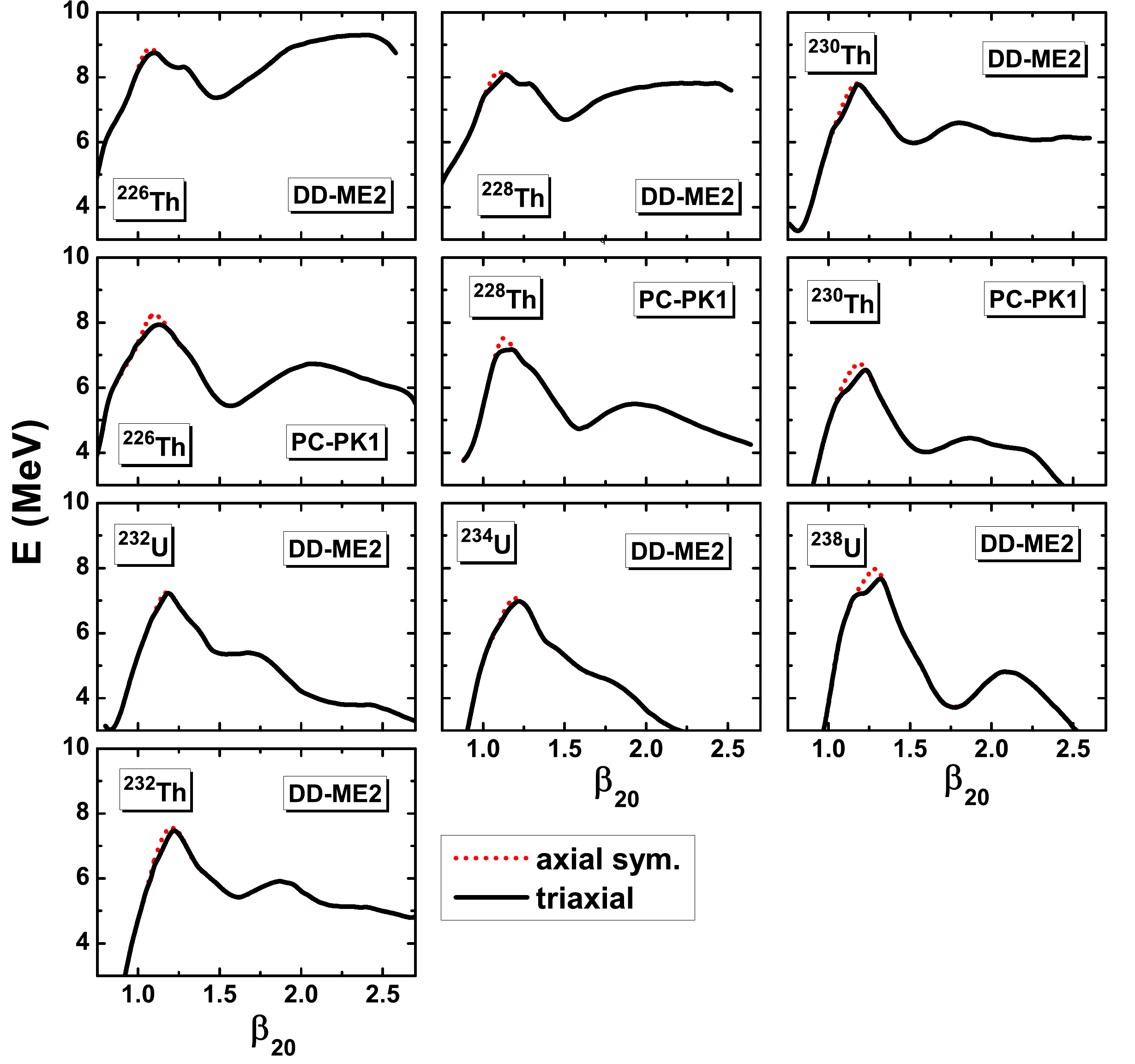}
}
\end{center}
\caption{(Color online)~\label{fig:Pathall}%
Potential energy curves [$E=E(\beta_{20})$] of $^{226,228,230,232}$Th and $^{232,234,238}$U
($E$ is normalized with respect to that of the ground state).
Taken from Ref.~\cite{Zhao2015_PRC91-014321}.
}
\end{figure*}

In Fig.~\ref{fig:pes}, we can find from the two-dimensional PES's 
calculated with effective interactions PC-PK1 or DD-ME2
that there appear third barriers and third minima in $^{226,228,230,232}$Th 
and $^{232,234,238}$U. 
Next we examine for these nuclei the one-dimensional potential energy curves 
which
are shown in Fig.~\ref{fig:Pathall}.
The excitation energy of the third minimum, 
the heights of the second and third fission barriers (relative to the ground state energy), 
and the depth of the third pocket defined as the difference between
the excitation energy of the third minimum and
the height of the third fission barrier, 
i.e., $\Delta E \equiv B_\mathrm{III} - E_\mathrm{III}$,
together with empirical parameters taken from Refs.~\cite{Capote2009_NDS110-3107,%
Blons1988_NPA477-231,Yoneama1996_NPA604-263,Csige2009_PRC80-011301R,%
Krasznahorkay1999_PLB461-15,Csige2013_PRC87-044321} are tabulated
in Table~\ref{tab:parameter}.

\begin{table}
\caption{\label{tab:parameter} %
Heights of the second and third saddle points, 
$B_{\rm II}$ and $B_{\rm III}$, 
and energy of the third minimum $E_{\rm III}$ (in MeV)
relative to the ground state for even-even Th and U isotopes.
$\Delta E \equiv B_\mathrm{III} - E_\mathrm{III}$ 
is defined as the depth of the third potential well.
The empirical values 
are taken from Refs.~\cite{Capote2009_NDS110-3107,%
Blons1988_NPA477-231,Yoneama1996_NPA604-263,Csige2009_PRC80-011301R,%
Krasznahorkay1999_PLB461-15,Csige2013_PRC87-044321}
and denoted by ``Emp''.
Taken from Ref.~\cite{Zhao2015_PRC91-014321}.
}
\begin{tabular}{llllllcc}
\hline
\hline
 Nucleus     & Parameters & $B_{\rm II}$ & $E_{\rm III}$ & $B_{\rm III}$ & $\Delta E$ \\ 
\hline
 $^{226}$Th  & DD-ME2     & 8.76         & $7.37$        & $9.31$        & $1.94$     \\
             & PC-PK1     & 7.94         & $5.44$        & $6.73$        & $1.29$     \\
 $^{228}$Th  & DD-ME2     & 8.16         & $6.69$        & $7.82$        & $1.13$     \\
             & PC-PK1     & 7.19         & $4.72$        & $5.50$        & $0.78$     \\
 $^{230}$Th  & DD-ME2     & 7.84         & $5.97$        & $6.60$        & $0.63$     \\
             & PC-PK1     & 6.56         & $4.01$        & $4.45$        & $0.44$     \\
             & Emp~\cite{Capote2009_NDS110-3107} 
                          & 6.80         &               &               &            \\
             & Emp~\cite{Blons1988_NPA477-231}   
                          & 5.75         & $5.55$        & $6.45$        & $0.90$     \\
 $^{232}$Th  & DD-ME2     & 7.53         & $5.42$        & $5.92$        & $0.50$     \\
             & Emp~\cite{Capote2009_NDS110-3107} 
                          & 6.70         &               &               &            \\
             & Emp~\cite{Yoneama1996_NPA604-263} 
                          & 6.50         & $5.40$        & $5.70$        & $0.30$      \\
 $^{232}$U   & DD-ME2     & 7.25         & \textemdash   & \textemdash   & \textemdash \\
             & Emp~\cite{Capote2009_NDS110-3107} 
                          & 5.40         &               &               &             \\
             & Emp~\cite{Csige2009_PRC80-011301R}
                          & 4.91         & $3.20$        & $6.02$        & $2.82$      \\
 $^{234}$U   & DD-ME2     & 7.01         & \textemdash   & \textemdash   & \textemdash \\
             & Emp~\cite{Capote2009_NDS110-3107} 
                          & 5.50         &               &               &            \\
             & Emp~\cite{Krasznahorkay1999_PLB461-15}
                          &              & $3.1 $        &               &            \\
 $^{238}$U   & DD-ME2     & 7.70         & $3.70$        & $4.81$        & $1.11$     \\
             & Emp~\cite{Capote2009_NDS110-3107} 
                          & 5.50         &               &               &            \\
             & Emp~\cite{Csige2013_PRC87-044321} 
                          & 5.6          & $3.6 $        & $5.6 $        & $2.0 $     \\
\hline
\hline
\end{tabular}
\end{table}

\begin{figure*}
\begin{center}
\resizebox{0.99\textwidth}{!}{%
 \includegraphics{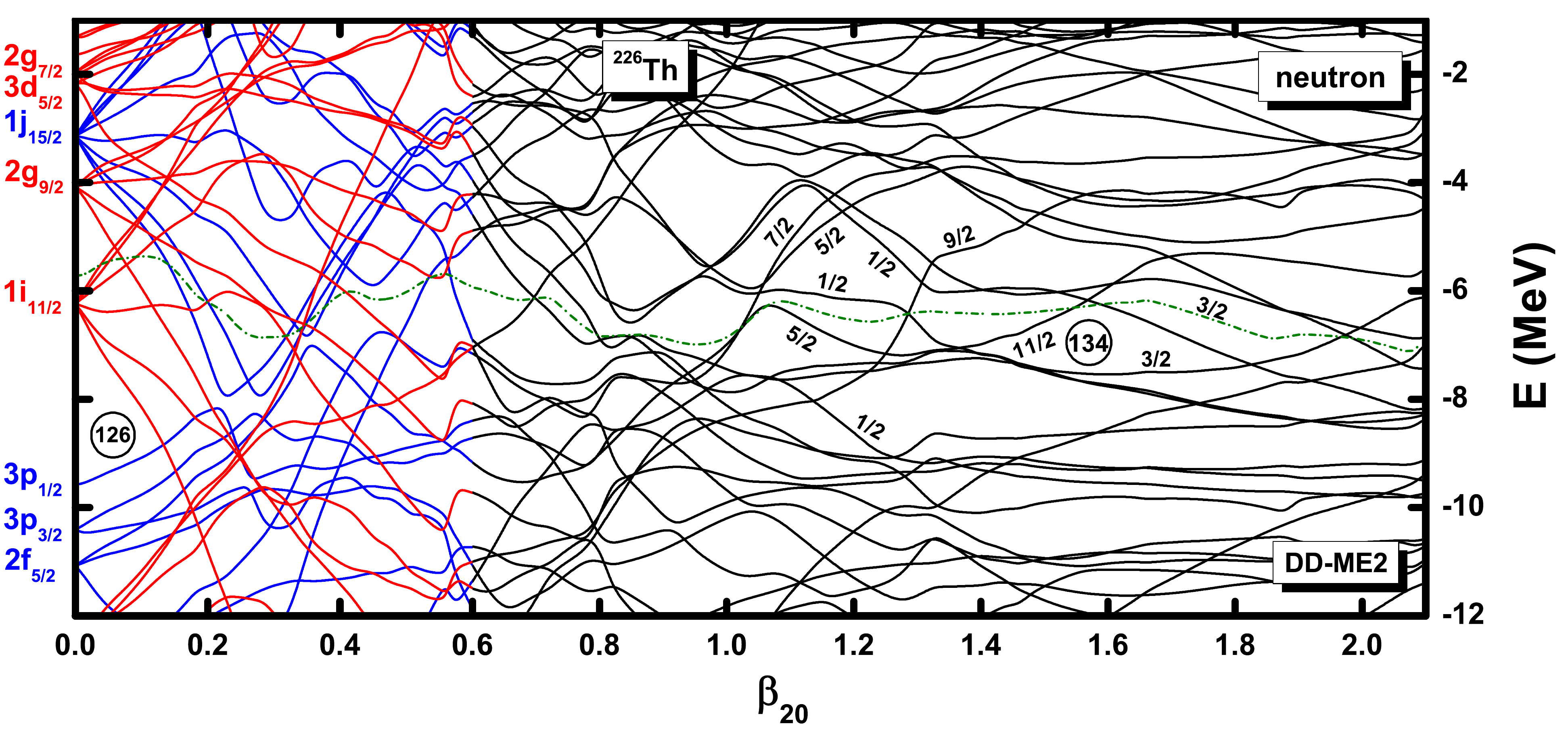}
}
\end{center}
\begin{center}
\resizebox{0.99\textwidth}{!}{%
 \includegraphics{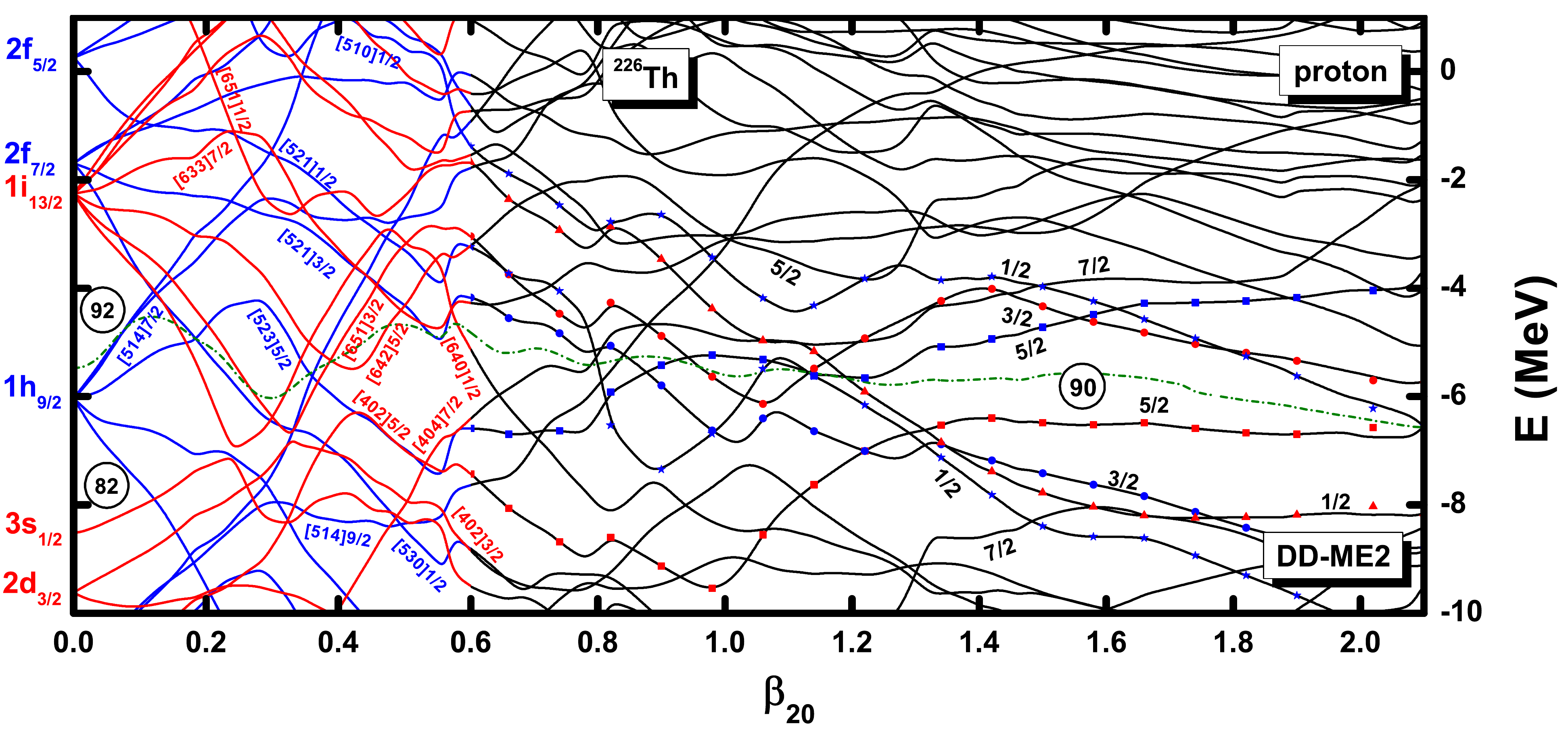}
}
\end{center}
\caption{(Color online)~\label{fig:lev}%
The single-particle level schemes for neutrons (upper panel) and protons (lower panel)
of $^{226}$Th near the Fermi surface along the static fission path.
For $\beta_{20} \leq 0.6$, only reflection-symmetric deformations are considered;
the red (blue) curves represent positive (negative) parity states and
the Nilsson quantum numbers are shown for some proton levels.
When $\beta_{20} > 0.6$, the octupole deformation $\beta_{30}$ has a non-vanishing
value, thus the parity $\pi$ is not a good quantum number and
only the projection of the single-particle angular momentum on the symmetry axis
is shown (the axial symmetry is assumed, see the text for more details).
The dash-dotted (green) curves denote the Fermi surface and
the red (blue) symbols in the lower panel are used to guide the eye.
The MDC-RMF calculation has been performed with the functional DD-ME2.
Taken from Ref.~\cite{Zhao2015_PRC91-014321}.
}
\end{figure*}

For $^{226}$Th, with effective interactions PC-PK1 and DD-ME2, the third minimum 
was predicted and the excitation energy is 5.44 and 7.37 MeV, respectively.
With PC-PK1, the second barrier is higher than the third one,
whereas the opposite is true with DD-ME2.
The depth of the third potential well calculated with PC-PK1 (DD-ME2) is 1.29 (1.94) MeV.
For $^{228}$Th, the third potential well becomes shallower
and the depth predicted from MDC-RMF calculations with PC-PK1 (DD-ME2) is 0.78 (1.13) MeV.
The third minimum is the most pronounced in $^{226,228}$Th 
among the nuclei considered in Ref.~\cite{Zhao2015_PRC91-014321}. 
For $^{230}$Th, $E_{\rm III}=5.55$ MeV and $B_{\rm III}=6.45$ MeV have been
deduced in Ref.~\cite{Blons1988_NPA477-231}.
The MDC-RMF model reproduced these values with DD-ME2,
but underestimated them by around 2 MeV with PC-PK1.
The calculation results with PC-PK1 and DD-ME2 are quite different for $^{232}$Th: 
There does not appear a third minimum in the PES calculated with PC-PK1; however,
the opposite is true for the case of DD-ME2.
A shallow third potential well with $\Delta E \approx 0.30$ MeV for $^{232}$Th
has been deduced in electron-induced fission cross section measurements 
\cite{Yoneama1996_NPA604-263}.
The MDC-RMF result with DD-ME2, $\Delta E = 0.50$ MeV, agrees reasonably well
with this empirical value.
In Table~\ref{tab:parameter}, one again finds that, with the neutron number $N$ increasing, 
the third potential pocket becomes shallower in the Th isotopes.

For U isotopes, the situation is much simpler than that for Th isotopes:
Only in $^{238}$U and with the effective interaction DD-ME2, 
the MDC-RMF calculations predict a third minimum.
As seen in Table~\ref{tab:parameter}, the calculated excitation energy of 
this minimum is 3.70 MeV and close to the empirical value 3.6 MeV \cite{Csige2013_PRC87-044321}.
But the calculated height of the third barrier is 4.81 MeV which is less than  
the empirical value 5.6 MeV \cite{Csige2013_PRC87-044321}.
Thus the MDC-RMF model prediction for the depth of the third potential
well in $^{238}$U, $\Delta E = 1.11$ MeV, is smaller than 
the corresponding empirical value 2.0 MeV \cite{Csige2013_PRC87-044321}.



\subsubsection{\label{sec:spl}Single-particle level structure}

The appearance of a hyperdeformed minimum roots in the
scheme of single-particle levels.
This has been discussed in Ref.~\cite{Zhao2015_PRC91-014321}.
Figure~\ref{fig:lev} shows the single-particle orbitals around  
the Fermi surface as functions of the quadrupole deformation $\beta_{20}$ for $^{226}$Th. 
Only the results with the DD-ME2 functional are shown.
The quadrupole deformation of the superdeformed state of $^{226}$Th locates
around $\beta_{20} \approx 0.6$.
For $\beta_{20} \leq 0.6$, the reflection symmetry is kept,
and thus the parity $\pi$ is conserved.
However, when $\beta_{20}>0.6$, the octupole deformation $\beta_{30}$ is non-zero and
the parity is not a good quantum number.
Furthermore, the triaxial deformation starts to be important around the second saddle point 
and the projection of the single-particle angular momentum on the symmetry axis 
is not a good quantum number either.
This could leads to a quite complicated single-particle structure 
around the second saddle point.
In Ref.~\cite{Zhao2015_PRC91-014321}, we were mainly focused on
single-particle orbitals beyond the second fission barrier,
we plotted single-particle levels obtained in axial MDC-RMF calculations
in Fig.~\ref{fig:lev}.

The third minimum of $^{226}$Th locates at 
$\beta_{20}\approx1.5$ and $\beta_{30}\approx0.7$.
From the neutron single-particle levels shown in the upper panel of Fig.~\ref{fig:lev}, 
one can find that around $\beta_{20}=1.5$, although the energy gap 
around the Fermi surface is not very large, the density of single-particle levels
is quite low.
As shown in the lower panel of Fig.~\ref{fig:lev}, 
there is a big energy gap around $Z=90$ in the region $\beta_{20} \approx 1.5$.
Thus it has been concluded in Ref.~\cite{Zhao2015_PRC91-014321} that 
the existence of the hyperdeformed minimum in the PES of $^{226}$Th 
is mainly due to the large proton gap at $Z=90$.
It has been noted that some other single-particle orbitals 
[dotted with red (blue) symbols and labeled with $\Omega$, i.e., the projection of
of the single-particle angular momentum on the symmetry axis 
in the lower panel of Fig.~\ref{fig:lev}] 
close to the proton Fermi surface are also responsible for
the formation of this energy gap.

\section{\label{sec:Y32}$Y_{32}$ correlations in $N=150$ isotones}

\begin{figure}
\begin{center}
\resizebox{0.45\textwidth}{!}{%
 \includegraphics{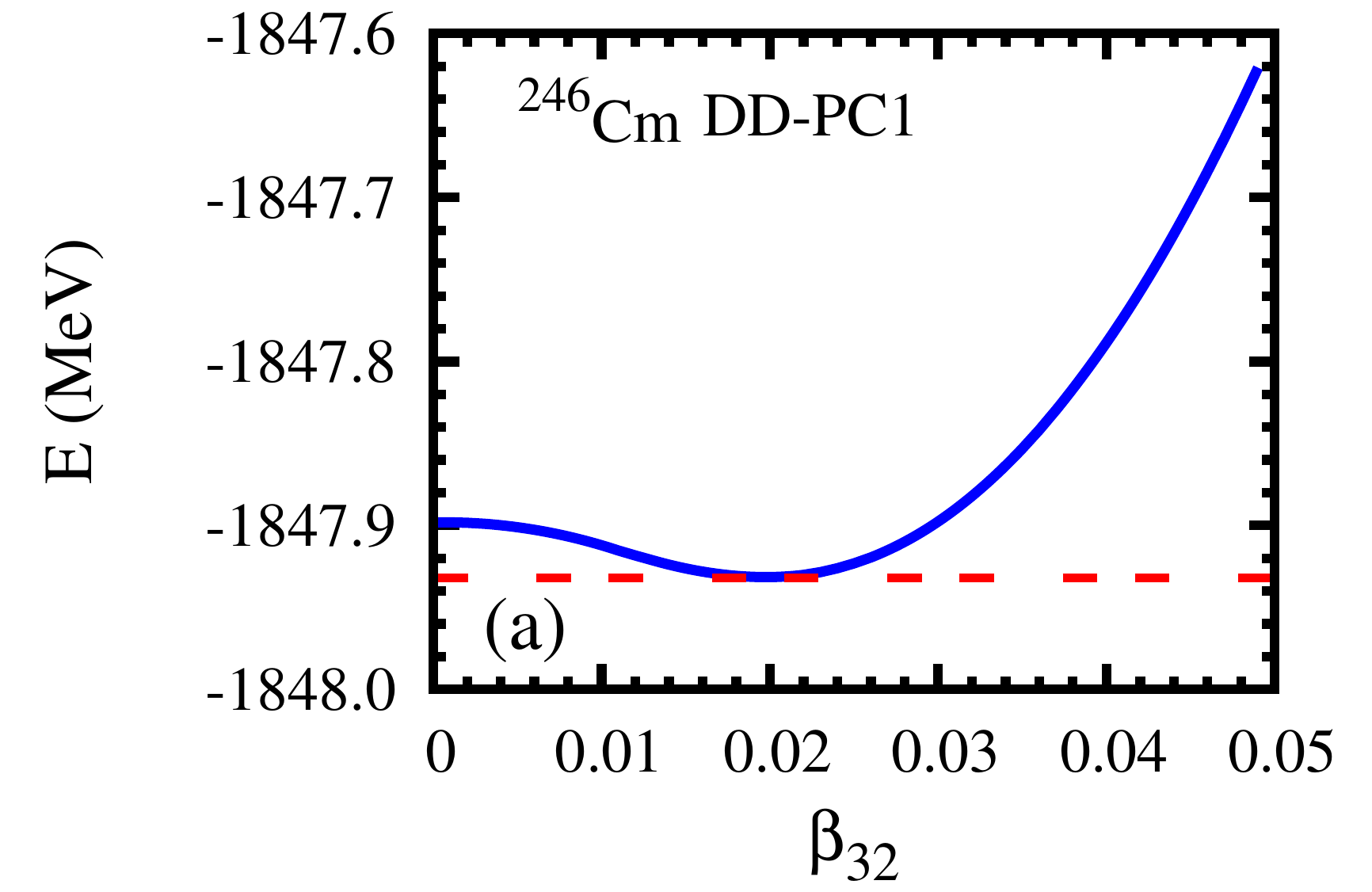} 
}
\end{center}
\begin{center}
\resizebox{0.45\textwidth}{!}{%
 \includegraphics{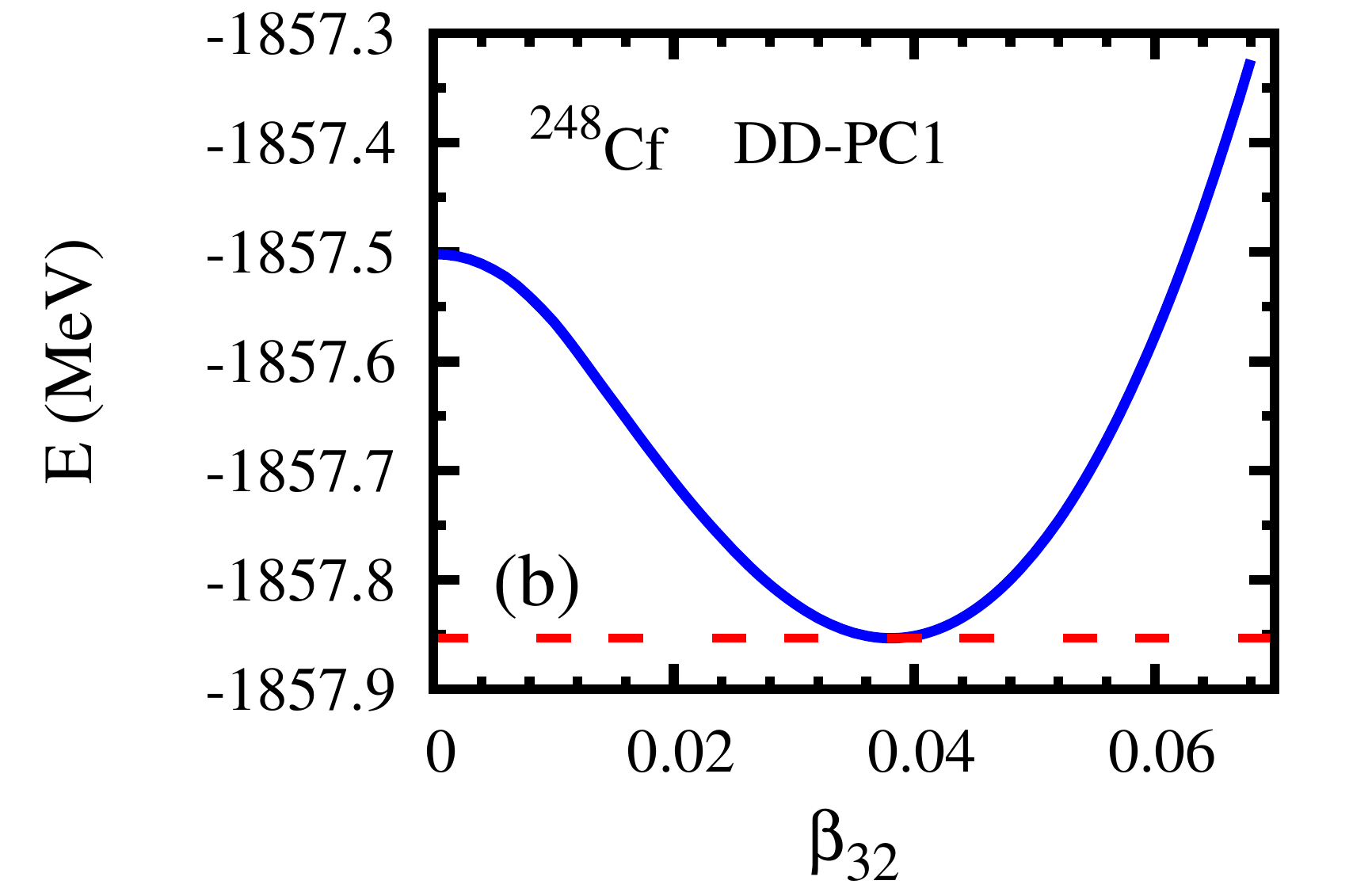} 
}
\end{center}
\begin{center}
\resizebox{0.45\textwidth}{!}{%
 \includegraphics{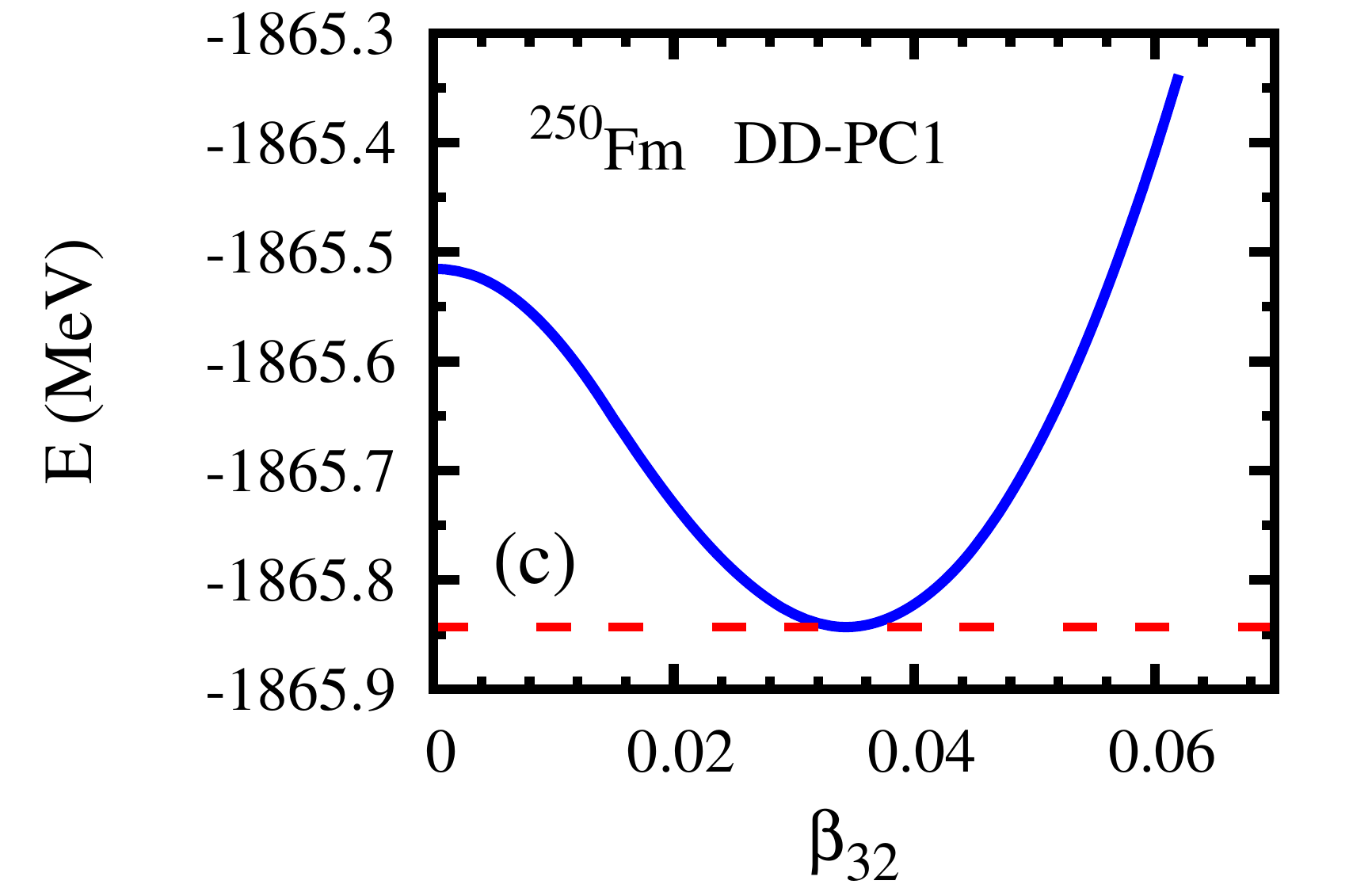} 
}
\end{center}
\begin{center}
\resizebox{0.45\textwidth}{!}{%
 \includegraphics{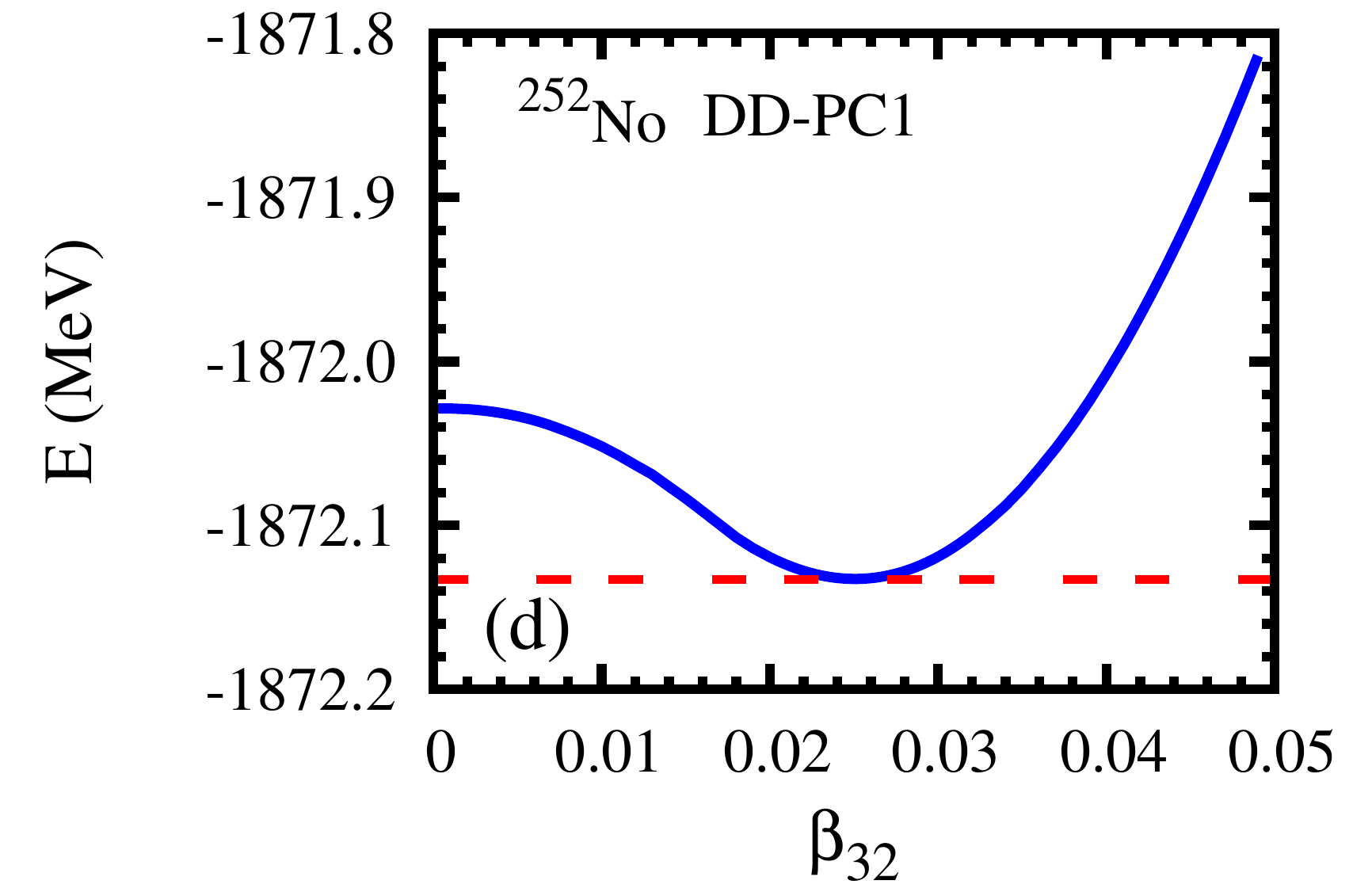} 
}
\end{center}
\caption{\label{fig:b32} (Color online)
Potential energy curves [$E=E(\beta_{32})$] for 
(a) $^{246}$Cm, (b) $^{248}$Cf, (c) $^{250}$Fm and
(d) $^{252}$No from MDC-RMF calculations
($E$ is normalized with respect to that of the ground state).
(b) is adapted from Ref.~\cite{Lu2014_PS89-054028}. 
}
\end{figure}

The pure non-axial octupole $\beta_{32}$ deformation (i.e., $\beta_{\lambda\mu}=0$ 
if $\lambda\ne3$ or $\mu\ne2$) has a tetrahedral symmetry.
The non-trivial irreducible representation of the symmetry group $T_d^D$
leads to highly degenerate single-particle levels and large shell gaps.
It has been predicted that the $\beta_{32}$ shape may appear
in the ground states of nuclei with certain combination of neutron and
proton numbers
\cite{Heiss1999_PRC60-034303,Li1994_PRC49-R1250,Dudek2002_PRL88-252502,Dudek2010_JPG37-064032}.
Recently, many theoretical efforts were devoted to the study of the tetrahedral shape
in atomic nuclei,
either from the $T_d^D$-symmetric single-particle level schemes
\cite{Li1994_PRC49-R1250,Dudek2002_PRL88-252502,Dudek2003_APPB34-2491,Dudek2007_IJMPE16-516} or
from different nuclear models including the macroscopic-microscopic
model \cite{Dudek2002_PRL88-252502,Dudek2007_IJMPE16-516,Schunck2004_PRC69-061305R,%
Dudek2006_PRL97-072501}, 
the Skyrme Hartree-Fock (SHF) model \cite{Dudek2007_IJMPE16-516,Schunck2004_PRC69-061305R,%
Yamagami2001_NPA693-579,Olbratowski2006_IJMPE15-333,Dudek2006_PRL97-072501,%
Zberecki2009_PRC79-014319,Takami1998_PLB431-242,Zberecki2006_PRC74-051302R},
and the reflection asymmetric shell model (RASM) 
\cite{Gao2004_CPL21-806,Chen2008_PRC77-061305R}.
A negative-parity band in $^{156}$Gd has been suggested to be
a candidate manifesting tetrahedral symmetry \cite{Dudek2006_PRL97-072501}.
Later, several experiments were made to investigate this feature in $^{156}$Gd
\cite{Bark2010_PRL104-022501,Jentschel2010_PRL104-222502}.
Although the existence of tetrahedral shape has not been confirmed experimentally, 
the interests in nuclear tetrahedral symmetry do stimulate more and more theoretical
studies related to point group symmetries in nuclei; the readers
are referred to Ref.~\cite{Dudek2013_APPB44-205} for a recent review on
this topic.

\begin{figure}[t]
\begin{center}
\resizebox{0.5\textwidth}{!}{%
 \includegraphics{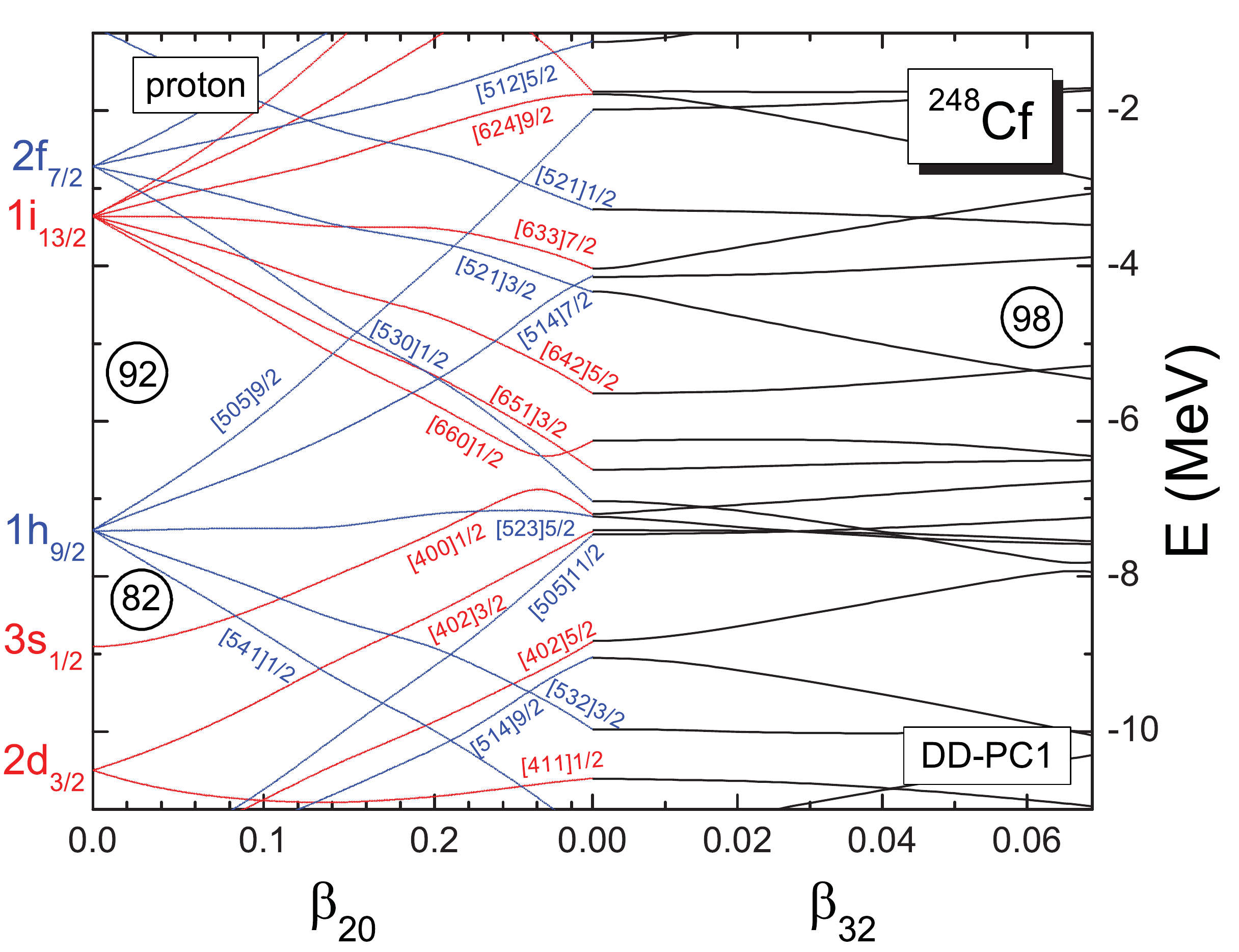}
}
\end{center}
\caption{\label{fig:lev-beta32_p}(Color online) %
The single-particle levels near the Fermi surface for protons of $^{248}$Cf 
as functions of quadrupole deformation
$\beta_{20}$ (left side) and of $\beta_{32}$ with $\beta_{20}$ fixed at 0.3 (right side).
Taken from Ref.~\cite{Zhao2012_PRC86-057304}.
}
\end{figure}

\begin{figure}[t]
\begin{center}
\resizebox{0.5\textwidth}{!}{%
 \includegraphics{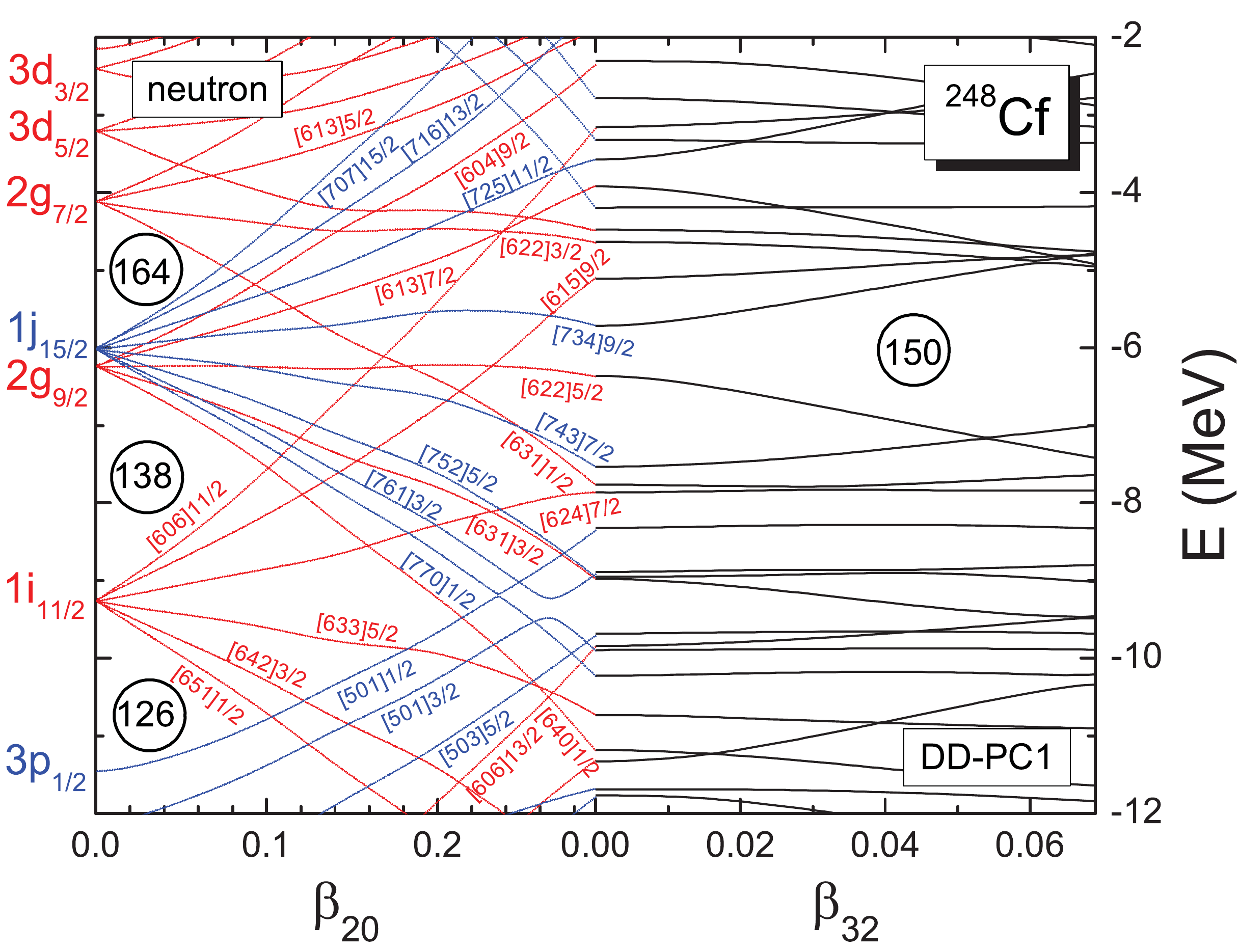}
}
\end{center}
\caption{\label{fig:lev-beta32_n}(Color online) %
The single-particle levels near the Fermi surface for neutrons 
of $^{248}$Cf as functions of quadrupole deformation
$\beta_{20}$ (left side) and of $\beta_{32}$ with $\beta_{20}$ fixed at 0.3 (right side).
Taken from Ref.~\cite{Zhao2012_PRC86-057304}.
}
\end{figure}

In recent years, the investigation of nuclei with $Z\sim 100$ becomes more and more
interesting 
because such studies can, on one hand, reveal the structure of these nuclei themselves
and on the other hand,
provide useful information for the structure of superheavy nuclei
\cite{Afanasjev2003_PRC67-024309,Leino2004_ARNPS54-175,Herzberg2006_Nature442-896,%
Herzberg2008_PPNP61-674,Zhang2011_PRC83-011304R,Zhang2012_PRC85-014324,Zhang2013_PRC87-054308}.
One interesting topic is about the low-lying $2^-$ states in several even-$Z$ $N = 150$ 
isotones \cite{Robinson2008_PRC78-034308}.
For example, in a quasiparticle phonon model with octupole correlations included,
the low-lying $2^-$ states in the isotones with $N = 150$
was explained by the existence of the octupole correlations originating 
from the neutron configuration $9/2^-[724] \otimes 5/2^+[622]$
and proton configurations $9/2^+[633] \otimes 5/2^-[521]$
or $7/2^+[633] \otimes 3/2^-[521]$ \cite{Jolos2011_JPG38-115103}.
Since in these two-quasiparticle configurations, the projections of the single-particle
angular momenta of like-quasiparticles on the symmetry axis ($K$) differ by 2,
$Y_{32}$ correlations should be more relevant than $Y_{30}$ correlations.
Indeed, Chen et al., using the reflection asymmetric shell model with $Y_{32}$ 
correlations considered, 
have reproduced quite well the low-energy $2^-$ rotational bands in the $N = 150$ isotones
\cite{Chen2008_PRC77-061305R}.
Furthermore, they also predicted that the strong non-axial octupole $Y_{32}$ effects
may still exist in superheavy nuclei with $Z=108$ \cite{Chen2008_PRC77-061305R} and
even heavier nuclei \cite{Chen2013_NPR30-278}.
Note that in the reflection asymmetric shell model and many other models, 
the values of $\beta_{32}$ are input parameters. 

We have carried out a self-consistent and microscopic study of
$Y_{32}$ effects in the $N = 150$ isotones by using 
the MDC-RMF model \cite{Zhao2012_PRC86-057304}.
It was found that $\beta_{32} > 0.03$ and 
the energy gain due to the inclusion of the $\beta_{32}$ deformation is 
more than 300 keV for the ground states of $^{248}$Cf and $^{250}$Fm. 
In $^{246}$Cm and $^{252}$No, there appear shallow $\beta_{32}$ minima compared
to $^{248}$Cf and $^{250}$Fm.
The origin of the occurrence of the $Y_{32}$ correlations was also
discussed.
In this Section, we will briefly present these results.

\subsection{Ground states and PECs around the minima}

We studied the ground states of nuclei $^{246}$Cm, $^{248}$Cf,
$^{250}$Fm and $^{252}$No by using the MDC-RMF model with the relativistic
density functional DD-PC1 \cite{Niksic2008_PRC78-034318}.
These four nuclei are all well deformed with 
$\beta_{20} \approx 0.3$ and $\beta_{40} \approx 0.1$.
But their $\beta_{22}$ and $\beta_{30}$ values are zero.
 
In Fig.~\ref{fig:b32}, the PECs [$E=E(\beta_{32})$] are shown
for $^{246}$Cm, $^{248}$Cf, $^{250}$Fm and $^{252}$No.
In the PECs, there exist clear potential pockets with 
the depth more than 0.3 MeV for $^{248}$Cf and $^{250}$Fm.
For $^{246}$Cm and $^{252}$No, only a shallow potential pocket appears 
along the $\beta_{32}$ shape degree of freedom.
%
For $^{246}$Cm, the ground state deformation $\beta_{32}\approx0.02$.
The PEC is rather flat around the minimum.
We have defined a quantity $E_\mathrm{depth}$, i.e., the energy difference 
between the ground state and the point with $\beta_{32}=0$,
to measure the energy gain due to the $\beta_{32}$ distortion.
For $^{246}$Cm, $E_\mathrm{depth} = 34$ keV, indicating a very small
$Y_{32}$ correlation.
For $^{248}$Cf, $^{250}$Fm and $^{252}$No, the potential minima 
appear $\beta_{32} \approx 0.037$, $0.034$ and $0.025$ 
and the energy gain $E_\mathrm{depth} = 0.351$, $0.328$ and $0.104$ MeV.
One can conclude that strong $Y_{32}$ correlations exist in these nuclei.

Both $\beta_{32}$ and $E_\mathrm{depth}$ reach maximal values at $^{248}$Cf 
in these four nuclei along the $N=150$ isotonic chain,
hinting that the $Y_{32}$ correlation is the strongest in $^{248}$Cf
\cite{Zhao2012_PRC86-057304}.
This conclusion is consistent with that given in
Refs.~\cite{Chen2008_PRC77-061305R,Jolos2011_JPG38-115103} and 
with the experimental fact 
that the $2^-$ state in $^{248}$Cf is the lowest one among those in these four
nuclei \cite{Robinson2008_PRC78-034308}.

\subsection{Single-particle level structure}

The non-axial octupole $Y_{32}$ correlation originates from the coupling 
between a pair of single-particle orbitals with $\Delta j = \Delta l = 3$ and $\Delta K=2$.
If such a pair of orbitals are nearly degenerate and near
the Fermi level, one can expect a strong $Y_{32}$ effect.
 
The single-particle states of $^{248}$Cf around the Fermi levels
are shown in Figs.~\ref{fig:lev-beta32_p} (for protons) and \ref{fig:lev-beta32_n} (for neutrons).
%
In Fig.~\ref{fig:lev-beta32_p}, it can be seen that
the two proton levels $\pi 2f_{7/2}$ and $\pi 1i_{13/2}$ are 
close to each other and this quasi degeneracy may result in octupole effects.
When $\beta_{20}$ increasing from 0 to 0.3, 
the two proton orbitals $[521]3/2$ and $[633]7/2$ 
keep nearly degenerate.
Since the quantum numbers of these two orbitals fulfill the conditions
for $Y_{32}$ correlations, i.e., 
$\Delta j = \Delta l = 3$ and $\Delta K=2$, the coupling 
between them leads to non-axial octupole $Y_{32}$ correlations.
With non-axial octupole deformation $\beta_{32}$ increasing from zero, 
an energy gap gradually develops at $Z=98$.
%
In Fig.~\ref{fig:lev-beta32_n}, one finds that 
the two neutron orbitals $[622]5/2$ (originating from $2g_{9/2}$) and 
$[734]9/2$ (originating from $1j_{15/2}$) are also nearly degenerate 
and just lie around the Fermi surface.
This results in an energy gap around $N=150$ 
with $\beta_{32}$ increasing.
Thus it was concluded that the $Y_{32}$ correlations 
are both from protons and from neutrons in these even-even $N=150$ isotones and 
such correlations are the most pronounced in $^{248}$Cf \cite{Zhao2012_PRC86-057304}.

From the above discussions and those made in Section~\ref{sec:spl},
it can be seen that a proper description of the single-particle structure 
is crucial for the appearance of the third barrier, the $\beta_{32}$ minimum, et al. 
Therefore a related issue becomes very crucial:
How well can covariant density functionals (or non-relativistic density functionals) 
describe single-particle level schemes in atomic nuclei? 
Nowadays this has become a hot topic; more details can be found in 
Refs.~\cite{Dobaczewski2015_NPA944-388,Afanasjev2015_JPG42-034002} and
references therein.

\subsection{Further discussions}

In Ref.~\cite{Zhao2012_PRC86-057304}, 
the dependence of the above conclusions on the functional form 
and on the effective interaction has been studied. Namely,
we investigated $^{246}$Cm, $^{248}$Cf, $^{250}$Fm and $^{252}$No 
with effective interactions PC-PK1 \cite{Zhao2010_PRC82-054319,Zhao2012_PRC86-064324},
DD-ME1 \cite{Niksic2002_PRC66-024306} and DD-ME2 
\cite{Lalazissis2005_PRC71-024312}.
The results with different parameter sets are similar and
several general conclusions were drawn:
(1) Density-dependent functionals give stronger $Y_{32}$
correlations than the nonlinear coupling ones;
(2) The non-axial octupole $Y_{32}$ effects in $^{246}$Cm are very weak;
(3) The $Y_{32}$ correlations are the strongest in $^{248}$Cf
among these four $N=150$ isotones 
except for that DD-ME1 gives a smaller energy gain for $^{248}$Cf 
than for $^{250}$Fm.

It is worthwhile to mention that the potential pocket in the $E\sim \beta_{32}$ curve 
is not quite deep for these four $N=150$ isotones. 
Therefore the existence of a minimum with non-zero $\beta_{32}$ 
in the $E\sim \beta_{32}$ curve does not necessarily mean that 
the corresponding nucleus has a static non-axial octupole shape.
But such a pocket could result in $Y_{32}$ correlations and in
lowering the energies of the $\beta_{32}$ vibrations.

\section{\label{sec:summary}Summary and perspectives}

Many shape degrees of freedom play crucial roles in the ground and low-lying states 
and fission properties of atomic nuclei.
It is important to include as many shape degrees of freedom as possible in a
self-consistent model for the study of nuclear shapes and potential energy surfaces.
This is particularly true for investigating atomic nuclei in unknown regions,
e.g., exotic nuclei and superheavy nuclei.

We have developed multidimensionally-constrained covariant density functional 
theories (MDC-CDFT)
\cite{Lu2012_PhD,Lu2012_PRC85-011301R,Lu2014_PRC89-014323,Zhao2016_in-prep}. 
In the MDC-CDFTs all shapes characterized by $\beta_{\lambda\mu}$ 
with even $\mu$ are considered.
The covariant density functional can be one of the following four forms: 
the meson exchange interaction or point-coupling nucleon interaction 
combined with 
the non-linear couplings or density-dependent couplings.
The pairing correlations are treated with the BCS approach (MDC-RMF) or 
the Bogoliubov transformation (MDC-RHB).  
The MDC-CDFTs have been applied to the study of 
potential energy surfaces and fission barriers of actinides
\cite{Lu2012_PRC85-011301R,Lu2012_EPJWoC38-05003,Lu2014_PRC89-014323,%
Lu2014_JPCS492-012014,Lu2014_PS89-054028,Zhao2015_PRC92-064315,Zhao2016_in-prep2}, 
the non-axial octupole $Y_{32}$ correlations in $N = 150$ isotones
\cite{Zhao2012_PRC86-057304},
the third minima and triple-humped barriers in light actinides \cite{Zhao2015_PRC91-014321}
and shapes of hypernuclei \cite{Lu2011_PRC84-014328,Lu2014_PRC89-044307}. 
In this Review, we presented the formalism and some applications
on normal nuclei.

In the study of fission barriers and potential energy surfaces of actinides, 
we found that 
besides the reflection asymmetric deformation, 
the non-axial deformation is also important for
the second fission barriers in actinide nuclei.
Both the outer and the inner barriers become lower when
the triaxial distortions are included in the MDC-RMF calculations. 
With the inclusion of the triaxial deformation, a good agreement 
between the calculated heights of outer fission barrier and
the empirical ones is achieved.

We explored possible existence of the third minima and the third barriers
in some light actinides and found that the appearance of the third minima
is dependent on the functionals used in the MDC-RMF calculations.
The functional DD-ME2 predicts more prominent third minima and barriers than PC-PK1.
A detailed analysis of the single-nucleon levels around the Fermi surface 
reveals that the formation of the third minimum mainly originates from the $Z=90$ 
proton energy gap at $\beta_{20} \approx 1.5$ and $\beta_{30} \approx 0.7$.

The non-axial reflection-asymmetric $\beta_{32}$ shape in $^{246}$Cm, $^{248}$Cf, 
$^{250}$Fm and $^{252}$No were investigated. The origin of the non-axial 
octupole $Y_{32}$ correlations in these nuclei was analyzed by examining 
the proton and neutron single-particle orbitals around the Fermi level. 
The $Y_{32}$ correlations originate from both protons and neutrons and 
such correlations are the most pronounced in $^{248}$Cf among
these four nuclei.

At present and in the future, we are carrying out or will perform the
following investigations:
\begin{itemize}
\item
One of our motivations to develop the MDC-CDFTs is to study shapes and
PES's of superheavy nuclei. 
We are performing MDC-CDFT calculations for superheavy nuclei
with the axial and reflection symmetries simultaneously broken.
\item
By constraining the shape to an extremely large deformation along a certain
shape degrees of freedom, one may force some $\alpha$-nuclei to be in 
a cluster configuration,
such as the linear chain shape obtained by constraining 
$\beta_2>0$ \cite{Yao2014_PRC90-054307,Zhao2015_PRL115-022501}, 
the tetrahedral shape by constraining $\beta_{32}$ 
\cite{Girod2013_PRL111-132503,Ebran2014_PRC89-031303R,%
Ebran2014_PRC90-054329}, 
the bubble shape by constraining the radius, 
the toroidal shape by constraining $\beta_2<0$, etc. 
The MDC-CDFTs provide a useful tool for the study of such {\it constraint cluster 
structure} in atomic nuclei. 
\item
Based on the MDC-CDFTs with one or more hyperon(s) included, 
we will study various shape effects in light and medium-heavy hypernuclei:
(1) Exotic shapes, e.g., superdeformed and hyperdeformed states, 
the tetrahedron-like state, et al.; 
(2) The interplay between localization (clustering) effects and spontaneous symmetry 
breaking of nuclear shapes;
(3) The effects of hyperon on fission properties;
(4) The dependence of the shape polarization effect of hyperon
and the persistence of superdeformed and hyperdeformed states on 
the effective interactions.
\item
Most of our investigations presented here have been performed for 
a limited number of nuclei and with only several selected effective interactions. 
Nowadays systematic studies of the global performance of many relativistic 
functionals on the ground state properties and beyond-mean-field correlations 
of nuclei across the nuclear chart have been performed, see, e.g., 
Refs.~\cite{Agbemava2014_PRC89-054320,Lu2015_PRC91-027304}. 
Such global analysis of the performance of various functionals 
on nuclear shapes, potential energy surfaces and fission barriers 
will be also carried out by using the MDC-CDFTs.
\end{itemize}


\begin{ack}
The author thanks
R. Capote, Y. S. Chen, 
S. N. Ershov, 
E. Hiyama, 
M. Isaka,
R. Jolos, 
M. Kimura, M. Kowal, 
H. Lenske, L. L. Li, H. Z. Liang, A. Lopez-Martens, B. N. Lu,
U.-G. Mei{\ss}ner, J. Meng, 
V. V. Pashkevich, 
P. Ring, 
H. Sagawa, N. Schunck, J. Skalski,
D. Vretenar, 
K. Wen, X. Z. Wu, 
J. M. Yao,
Z. H. Zhang,
E. G. Zhao,
J. Zhao, 
and 
B. S. Zou
for helpful discussions and/or fruitful collaborations 
on the work presented in this Review.
This work has been supported by 
the National Key Basic Research Program of China (Grant No. 2013CB834400), 
the National Natural Science Foundation of China (Grants No. 11120101005, 
No. 11275248 and No. 11525524),
and
the Knowledge Innovation Project of the Chinese Academy of Sciences 
(Grant No. KJCX2-EW-N01). 
The results described in this paper are obtained on 
the High-performance Computing Cluster of SKLTP/ITP-CAS and
the ScGrid of the Supercomputing Center, Computer Network Information Center of 
the Chinese Academy of Sciences.
\end{ack}



\providecommand{\newblock}{}

\end{document}